\documentclass[sigconf,9pt]{acmart}

\usepackage[english]{babel}
\usepackage{blindtext}

\usepackage{color}
\usepackage{multirow}
\usepackage{url}
\usepackage{graphicx}
\usepackage{algorithmicx,algorithm}
\usepackage[noend]{algpseudocode}

\usepackage{enumitem} 
\usepackage{latexsym}
\usepackage{amsmath}
\usepackage{amsfonts}
\usepackage{tabularx}
\usepackage{balance}
\usepackage{subcaption}
\usepackage{caption}
\usepackage{enumitem} 
\usepackage{xspace}
\usepackage{booktabs}
\usepackage{tabularx}
\usepackage{balance}
\usepackage{graphicx}
\usepackage{svg}
\usepackage{pifont}
\usepackage[font=small]{caption}
\svgsetup{
    inkscapepath=i/svg-inkscape/
}
\svgpath{{svg/}}
\usepackage{etoolbox}
\usepackage[all]{nowidow}

\setlist{nolistsep}

\newcommand{\parabf}[1]{\medskip\noindent\textbf{#1}}
\newcommand{\parait}[1]{\medskip\noindent\textit{#1}}
\newcommand{\paraf}[1]{\noindent\textbf{#1}}
\newcommand{\cut}[1]{}

\newcommand{\sysname}{DistTrain\xspace}

\newcommand{\revision}[1]{\textcolor{black}{#1}}
\newcommand{\revisionsigcomm}[1]{\textcolor{black}{#1}}

\usepackage{array}

\ccsdesc[500]{Computer systems organization~Cloud computing}
\ccsdesc[500]{Computing methodologies~Machine learning}

\keywords{Large Language Models, Multimodal Models, Distributed Training}

\setcopyright{acmlicensed}
\acmYear{2025}\copyrightyear{2025}
\acmConference[SIGCOMM '25]{ACM SIGCOMM 2025 Conference}{September 8--11, 2025}{Coimbra, Portugal}
\acmBooktitle{ACM SIGCOMM 2025 Conference (SIGCOMM '25), September 8--11, 2025, Coimbra, Portugal}
\acmDOI{10.1145/3718958.3750472}
\acmISBN{979-8-4007-1524-2/25/09}

\acmPrice{}

\definecolor{ao}{rgb}{0.0, 0.5, 0.0}

\begin{document}
\sloppy
\date{}
\title{
    DistTrain: Addressing Model and Data Heterogeneity with Disaggregated Training for Multimodal Large Language Models}

\author{
    \vspace{-0.0in}
    Zili Zhang$^{\ast}$\qquad Yinmin Zhong$^{\ast}$\qquad Yimin Jiang$^{\star}$\qquad Hanpeng Hu$^{\dagger}$\qquad \\
    \vspace{0.1in}
    Jianjian Sun$^{\dagger}$\qquad Zheng Ge$^{\dagger}$\qquad Yibo Zhu$^{\dagger}$\qquad Daxin Jiang$^{\dagger}$\qquad Xin Jin$^{\ast}$\\
    \vspace{0.12in}
    $^{\ast}$\textit{School of Computer Science, Peking University} \qquad $^{\dagger}$\textit{StepFun} \qquad $^{\star}$\textit{Independent Researcher}\\
    \vspace{0.2in}
} 

\renewcommand{\shortauthors}{Z. Zhang et al.}
\begin{sloppypar}
\begin{abstract}
Multimodal large language models (LLMs) empower LLMs to ingest inputs and
generate outputs in multiple forms, such as text, image, and audio. However, the
integration of multiple modalities introduces heterogeneity in both the model and
training data, creating unique systems challenges.

We propose \sysname, a disaggregated training system for multimodal LLMs.
\sysname incorporates two novel disaggregation techniques to address model and
data heterogeneity, respectively. The first is \emph{disaggregated model
orchestration}, which separates the training for modality encoder, LLM
backbone, and modality generator. This allows the three components to adaptively
and independently orchestrate their resources and parallelism configurations.
The second is \emph{disaggregated data preprocessing}, which decouples data preprocessing from
training. 
This eliminates resource contention between preprocessing and
training, and enables efficient data reordering to mitigate stragglers within
and between microbatches caused by data heterogeneity.
We evaluate \sysname across different sizes of multimodal LLMs on a
large-scale production cluster. The experimental results
show that \sysname achieves 54.7\% Model FLOPs Utilization (MFU) when training a
72B multimodal LLM on 1172 GPUs and outperforms Megatron-LM by up to 2.2$\times$
on training throughput.
\end{abstract}

\maketitle

\vspace{-6pt}
{\par\smallskip\small\noindent{\bfseries ACM Reference Format:}\par\nobreak
  \noindent Zili Zhang, Yinmin Zhong, Yimin Jiang, Hanpeng Hu, Jianjian Sun, Zheng Ge, Yibo Zhu, Daxin Jiang, Xin Jin. 2025. DistTrain: Addressing Model and Data Heterogeneity with Disaggregated Training for Multimodal Large Language Models.
  In \textit{SIGCOMM '25, August 8-August 11, 2025, Coimbra, Portugal, 15 pages.}
}

\title{DistTrain: Addressing Model and Data Heterogeneity with Disaggregated Training ...}
\section{Introduction}
\label{sec:introduction}

\revision{
Traditional multimodal models, which focus solely on either cross-modal
feature representation (e.g., CLIP~\cite{radford2021learning}) or multimodal content generation (e.g., Stable-Diffusion~\cite{rombach2022high} and \revisionsigcomm{VAE~\cite{kingma2013auto}}),
cannot simultaneously process multimodal inputs and generate multimodal outputs.
This restricts their capabilities and hinders seamless human interaction.
Multimodal large language models (LLMs) emerge to integrate the strengths of
LLMs~\cite{vaswani2017attention, brown2020language, touvron2023llama}
with traditional multimodal models to simultaneously ingest multimodal inputs and
generate multimodal outputs in auto-regressive manner.
Multimodal
LLMs have already shown great potential in tasks such as vision
understanding and generation~\cite{liu2024world, zhang2023internlm, wang2024visionllm,
dong2023dreamllm, kondratyuk2023videopoet}, audio comprehension~\cite{rubenstein2023audiopalm,
borsos2023audiolm}, and embodied AI~\cite{zhang2023large, driess2023palm}.
Many organizations are actively developing multimodal LLMs, such as OpenAI's
GPT-4o~\cite{gpt4-o}, Google's Gemini~\cite{gemini}, and Meta's Llama3~\cite{llama}, etc.
}

\begin{figure}[t!]
    \centering
    \includegraphics[width=0.93\linewidth]{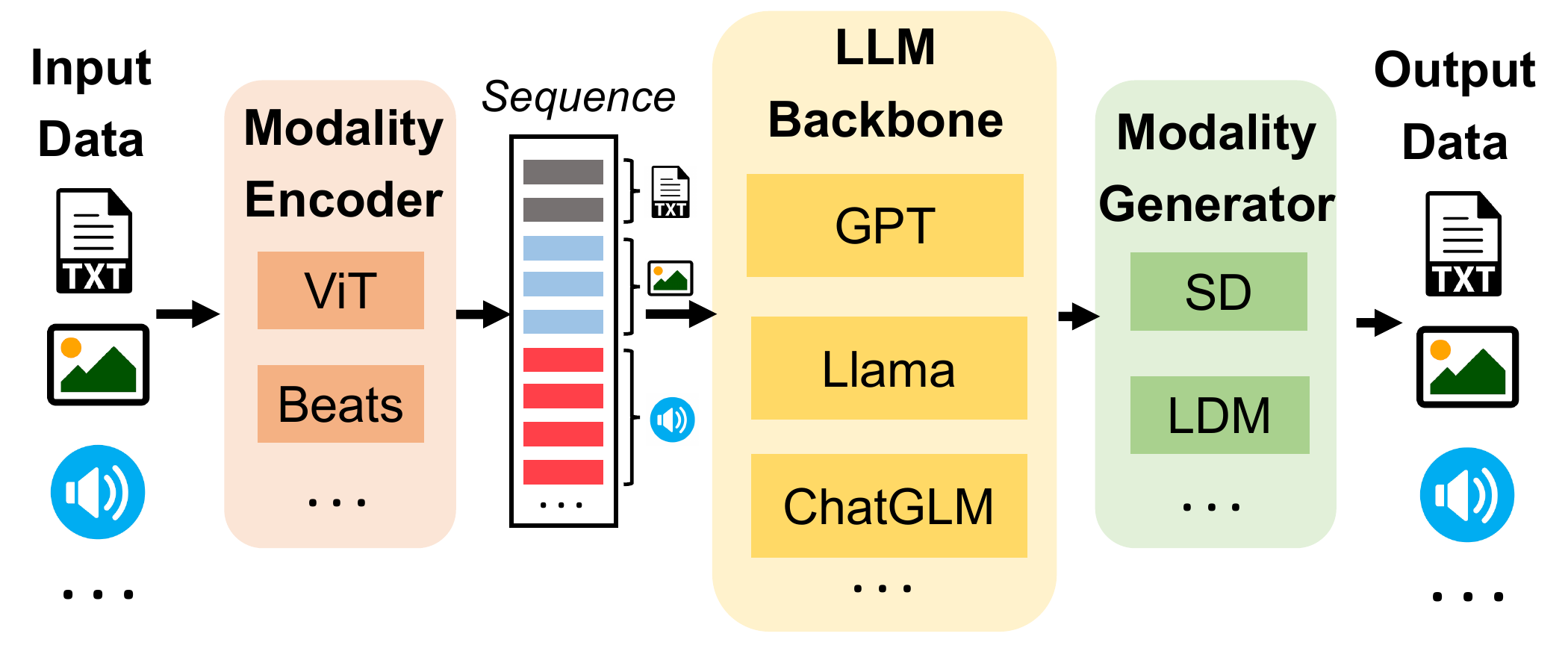}
    \vspace{-0.19in}
    \caption{\revision{The architecture of multimodal LLMs}.}
    \vspace{-0.17in}
    \label{fig:intro:model}
\end{figure}

Figure~\ref{fig:intro:model} depicts the model architecture of
multimodal LLMs, comprising three modules: modality encoder, LLM backbone, and
modality generator~\cite{dong2023dreamllm, yin2023survey, zhang2024mm}.
\revision{We emphasize this architecture underpins most state-of-the-art multimodal LLMs (Table~\ref{bg::MLLM-arch}).
}
These modules are linked
by projectors, which may incorporate MLP or cross-attention.
\revision{ The modality encoder transforms inputs from various modalities into different
modality tokens. These tokens from different modality are interleaved as a \emph{sequence} for LLM training.}
The modality
generator translates the processed information back into coherent outputs
tailored to each modality.

Training multimodal LLMs is more challenging than training traditional multimodal
models. Traditional
multimodal model training frameworks like DistMM~\cite{huang2024distmm} lack support
for simultaneous training of encoders, LLMs, and generators, particularly techniques tailored for LLMs.
The current practice for training multimodal LLMs~\cite{alayrac2022flamingo, wu2024deepseek} extends LLM
training frameworks (e.g., Megatron-LM~\cite{shoeybi2019megatron}) with
additional modality encoder and generator modules,
which uses a monolithic approach. In
this context, the monolithic training refers to two aspects. First, it
uses the same data and tensor parallelism strategies of distributed training across the modality
encoder, LLM backbone, and modality generator modules. As for pipeline parallelism, the
encoder and generator modules are incorporated as additional pipeline stages in the
training pipeline.
Second, it couples data preprocessing into training process,
and co-locates data preprocessing and training on the same machines.

The monolithic approach falls short of handling unique systems challenges
in multimodal LLM training. Compared to traditional LLM training that
only handles text, the multiple modalities in multimodal LLM training introduces
\emph{model heterogeneity} and \emph{data heterogeneity}. Model heterogeneity
stems from processing multiple modalities with different modules (i.e.,
encoder, LLM backbone, and generator) that vary dramatically in size
and operator complexity. Using the same parallelism strategies across different
modules introduces severe pipeline bubbles, resulting in poor GPU utilization (\S\ref{sec:background:model}).
Data heterogeneity emerges from processing multimodal data that differ
significantly in volume and processing cost, e.g., a high-resolution image is
much larger and requires more resources for preprocessing than a line of text.
In the monolithic approach, where data preprocessing and training are integrated
and co-located, the substantial processing cost significantly hinders the training process.
Additionally, the heterogeneity across different modality data (i.e., each training
sample has a varying number of modality tokens) leads to inter-microbatch and
intra-microbatch training stragglers, which prolong the training duration and
exacerbate the pipeline bubbles (\S\ref{sec:background:data}).
In our initial efforts using the monolithic approach,
we observed extremely low MFU in production training, as low as 20\% (\S\ref{sec:evaluation:overall}).

To this end, we propose \sysname, a disaggregated training system for multimodal
LLMs. \sysname applies the systems principle of disaggregation in the context of
multimodal LLM training to address the unique systems challenges. We design two novel disaggregation techniques, which are
\emph{disaggregated model orchestration} and \emph{disaggregated data
preprocessing}, to address model and data heterogeneity, respectively.

For model heterogeneity, we analyze the pipeline bubbles caused by multiple
modalities across different modules in multimodal LLMs, and identify their root causes
(\S\ref{sec:background:model}). To address this, we design \emph{disaggregated model
orchestration} to separate the training of modality encoder, LLM
backbone, and modality generator. This separation enables adaptive orchestration
of resources and parallelism configurations across the three modules. Building
on this, we formulate the model orchestration as an optimization problem aimed at
minimizing the training time per iteration. We then design an adaptive model orchestration
algorithm that navigates the complicated search space to find the \emph{optimal}
resource and parallelism configurations. Disaggregated model orchestration is able to
minimize pipeline bubbles caused by model heterogeneity and achieves \emph{optimal}
training efficiency.

For data heterogeneity, we characterize the training data of different modalities,
and unveil their preprocessing overheads and the stragglers they brought into the training process (\S\ref{sec:background:data}). We
categorize the stragglers into \emph{intra-microbatch} and
\emph{inter-microbatch} stragglers. We design
\emph{disaggregated data preprocessing}
to eliminate the interference between preprocessing and training.
Importantly, such disaggregation enables flexible
data preprocessing, and provides new opportunities for \emph{data reordering} to
mitigate stragglers without additional overhead. We use intra-microbatch reordering to evenly
distribute the load across data parallelism groups, and inter-microbatch
reordering to hide the training stragglers into the training pipeline.
\revisionsigcomm{
This dual-level reordering effectively mitigates the stragglers caused by data heterogeneity.
Importantly, since the reordering only permutes the sequence of commutative gradient accumulation operations,
it preserves the training's convergence semantics.
}

In summary, we make the following contributions.
\begin{itemize}[leftmargin=*]
    \item We identify the unique system challenges introduced by multiple
    modalities in multimodal LLM training, and summarize them as model
    and data heterogeneity.

    \item We propose \sysname, a disaggregated training system for multimodal
    LLMs, and design two novel techniques---disaggregated model orchestration
    and disaggregated data preprocessing---to address model and data
    heterogeneity.

    \item We implement \sysname and conduct experiments on our production
    cluster with thousands of GPUs. The experimental results show that \sysname
    achieves 54.7\% MFU when training a 72B multimodal LLM on 1172 GPUs and
    outperforms Megatron-LM by up to 2.2$\times$ on training throughput.
\end{itemize}
\section{Motivation}
\label{sec:background}

\subsection{Multimodal LLM Training}
\label{sec:background:bg}

\paraf{Traditional multimodal models.}
Traditional multimodal models encompass diverse architectures,
such as contrastive learning~\cite{radford2021learning} and stable diffusion~\cite{rombach2022high} (SD).
Contrastive learning uses encoder models to capture cross-modal similarities, primarily for feature representation.
Stable diffusion generates images from text prompts by iteratively refining random noise through a denoising process, enabling text-to-image generation.
However, these models are task-specific and lack the flexibility to handle diverse multimodal scenarios,
which cannot process multimodal inputs and generate multimodal outputs simultaneously,
restricting their model capabilities in complex real-world settings and hindering seamless interaction with humans.

\begin{table}[t!]
    \centering
    \resizebox{\linewidth}{!} {
    \begin{tabular}{cccc}
        \toprule
        \textbf{Multimodal} & \multirow{2}{*}{\textbf{Encoder(s)}} & \textbf{LLM} & \multirow{2}{*}{\textbf{Generator(s)}} \\
        \textbf{LLM} & & \textbf{Backbone} & \\
        \midrule
        Flamingo~\cite{alayrac2022flamingo} & NFNet~\cite{brock2021high} & GPT-3~\cite{brown2020language} & LM-Head \\
        LLaVA~\cite{liu2024visual} & CLIP~\cite{radford2021learning} & Vicuna~\cite{chiang2023vicuna} & LM-Head \\
        PaLM-E~\cite{driess2023palm} & ViT~\cite{dosovitskiy2020image} & PaLM~\cite{chowdhery2023palm} & LM-Head \\
        EMU~\cite{sun2023emu} & EVA-CLIP~\cite{sun2023eva} & Llama~\cite{touvron2023llama} & LM-Head, SD~\cite{rombach2022high} \\
        Bagel~\cite{deng2025emerging} & ViT~\cite{sun2023eva} & Qwen2.5~\cite{yang2025qwen} & LM-Head, VAE~\cite{kingma2013auto} \\
        \multirow{2}{*}{VideoPoet~\cite{kondratyuk2023videopoet}} & MAGViT~\cite{yu2023language}, & \multirow{2}{*}{GPT~\cite{radford2018improving}} & MAGViT~\cite{yu2023language}, \\
        & SoundStream~\cite{zeghidour2021soundstream} & & SoundStream~\cite{zeghidour2021soundstream} \\
        \bottomrule
    \end{tabular}
    }
    \vspace{0.0in}
    \caption{\revisionsigcomm{Some examples of multimodal LLM's architecture.}}
    \vspace{-0.25in}
    \label{bg::MLLM-arch}
\end{table}

\parabf{Multimodal LLMs.}
Large language models (LLMs)\cite{achiam2023gpt, gemini, llama},
built on homogeneous transformer layers and trained on extensive text corpora,
excel at understanding and generating high-quality text but are limited to
text-only processing. 
Multimodal LLMs address these limitations by integrating the strengths of LLMs
with traditional multimodal models,
enabling the simultaneous processing of diverse inputs
(e.g., images, audio, video) and generating multimodal outputs in auto-regressive manner.
For example, GPT-4o~\cite{gpt4-o} gains wide attention by facilitating more natural
interaction with humans through both visual and auditory modalities.


Figure~\ref{fig:intro:model} illustrates the model architecture of a multimodal
LLM, which consists of three modules: a modality encoder, an LLM backbone, and a
modality generator~\cite{alayrac2022flamingo, yin2023survey, zhang2024mm}.
Different from traditional multimodal models, this generative model facilitates a more robust and interpretable process for
understanding and generating multimodal contents through auto-regressive decoding.
Specifically, the modality encoder transforms input data from different modalities (e.g.,
ViT~\cite{dosovitskiy2020image} for images and Beats~\cite{chen2022beats} for
audios) into an intermediate representation, which
is then converted into different modality tokens with input
projector (e.g., MLP and cross-attention).
These tokens from different modalities form a \emph{sequence} for LLM training.
The LLM backbone, typically a
transformer model (e.g., GPT~\cite{radford2019language, brown2020language} and
Llama~\cite{llama}), processes the input \emph{sequence} to extract
intricate data patterns and inter-modal relationships.
The LLM backbone's output is refined by an output projector and then passed to a modality generator (e.g., Diffusion~\cite{rombach2022high}
for images, AudioLDM~\cite{liu2023audioldm} for audio), which converts the processed information into the corresponding modal outputs.
Table~\ref{bg::MLLM-arch} summarizes the three modules of some state-of-the-art multimodal LLMs.
\revisionsigcomm{
We omit text tokenizer (e.g., BPE) and encoder (e.g., text embedding) from the table's encoder column.
LM-Head is responsible for decoding the text tokens
}

\revision{
Multimodal LLM training necessitates training all three modules simultaneously.
Additionally, during different training phases, specific modules are frozen to stabilize training loss~\cite{chen2025janus, deng2025emerging}.
Regarding the data, input sequences comprise text, image, and audio tokens.
The data from different modalities are encoded into \emph{subsequences} which are then interleaved to form fixed-length training \emph{sequences}~\cite{team2024chameleon}.
}

\begin{figure}[t!]
    \centering
    \includegraphics[width=0.98\linewidth]{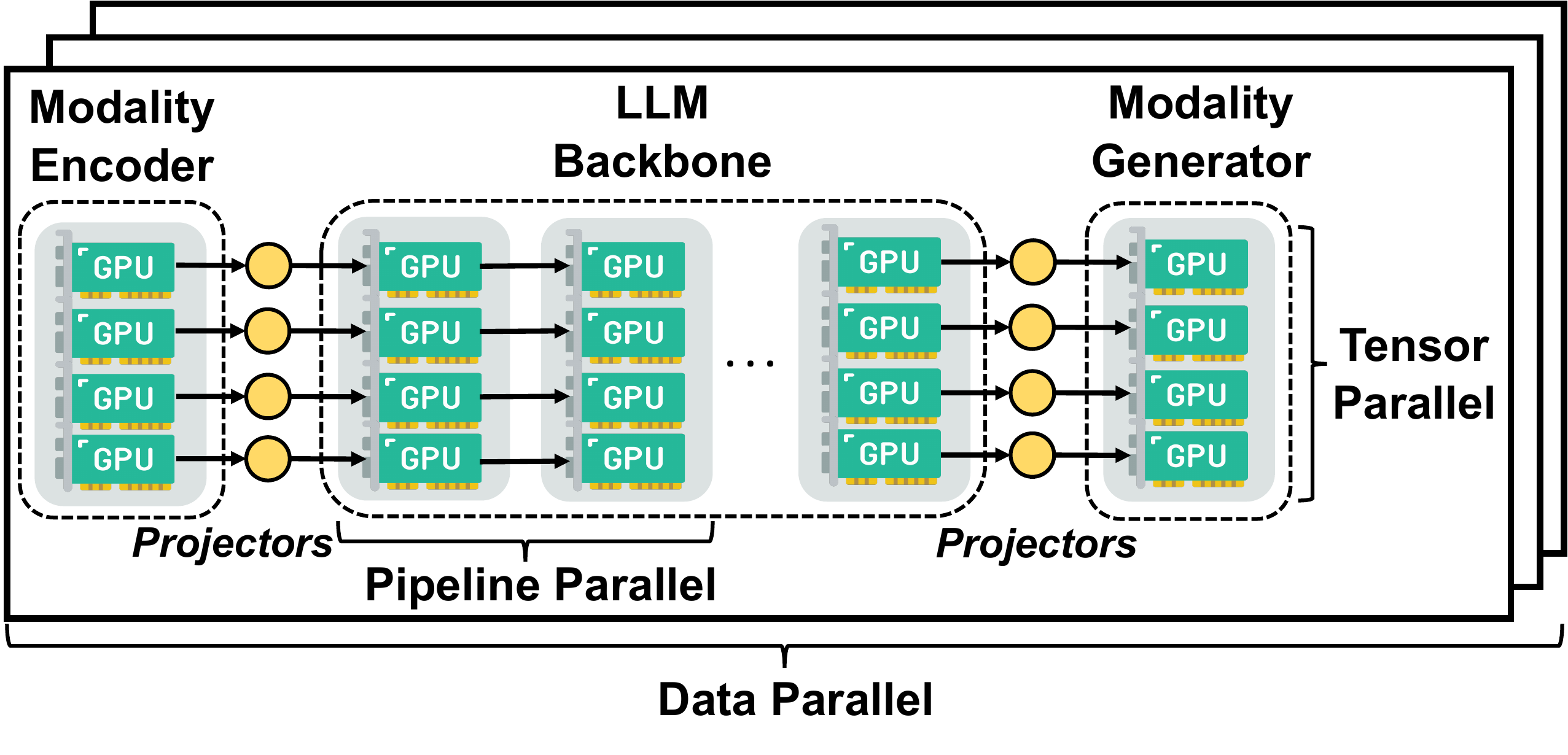}
    \vspace{-0.17in}
    \caption{Training multimodal LLMs with Megatron-LM.}
    \vspace{-0.1in}
    \label{fig:background:megatron_training}
\end{figure}

\parabf{Training frameworks.}
\revision{
Multimodal model training frameworks such as DistMM~\cite{huang2024distmm} do not support
training encoder, LLM backbone and generator simultaneously.
Especially, they lack support training techniques~\cite{chen2024centauri, narayanan2021efficient, rajbhandari2020zero} for LLMs.
In contrast, LLM training frameworks like Megatron-LM~\cite{shoeybi2019megatron} can be
extended with modality encoder and generator modules,
making them widely used in production multimodal LLM training, e.g., Flamingo~\cite{alayrac2022flamingo}, DeepSeek-VL2~\cite{wu2024deepseek}
and VideoPoet~\cite{kondratyuk2023videopoet}.
Figure~\ref{fig:background:megatron_training} shows training multimodal LLMs
with Megatron-LM. Megatron-LM adds multimodal modules as additional
layers, and incorporates additional pipeline parallelism (PP)
stages to accommodate encoder and generator. The same tensor
parallelism (TP) strategy used in the LLM backbone is applied to encoder and generator.
If the encoder and generator are not large enough,
they are replicated across the GPUs in the TP group to maximize resource
utilization. As for data parallelism (DP), Megatron-LM applies the same DP
strategy to the multimodal modules as the LLM backbone. The projector is
co-located with the encoder and generator, and is replicated across the
GPUs in the TP group. Moreover, data preprocessing is incorporated into the
training process and co-located with training on the same node. This approach is
\emph{monolithic} in terms that the encoder and generator use
the same parallelism strategy as the LLM backbone, and that preprocessing and
training are integrated and co-located. Such a monolithic approach introduces significant
computation imbalance stemming from \emph{model} and \emph{data} heterogeneity.
}

\subsection{Model Heterogeneity}
\label{sec:background:model}

\begin{figure}[t!]
    \centering
    \includegraphics[width=0.82\linewidth]{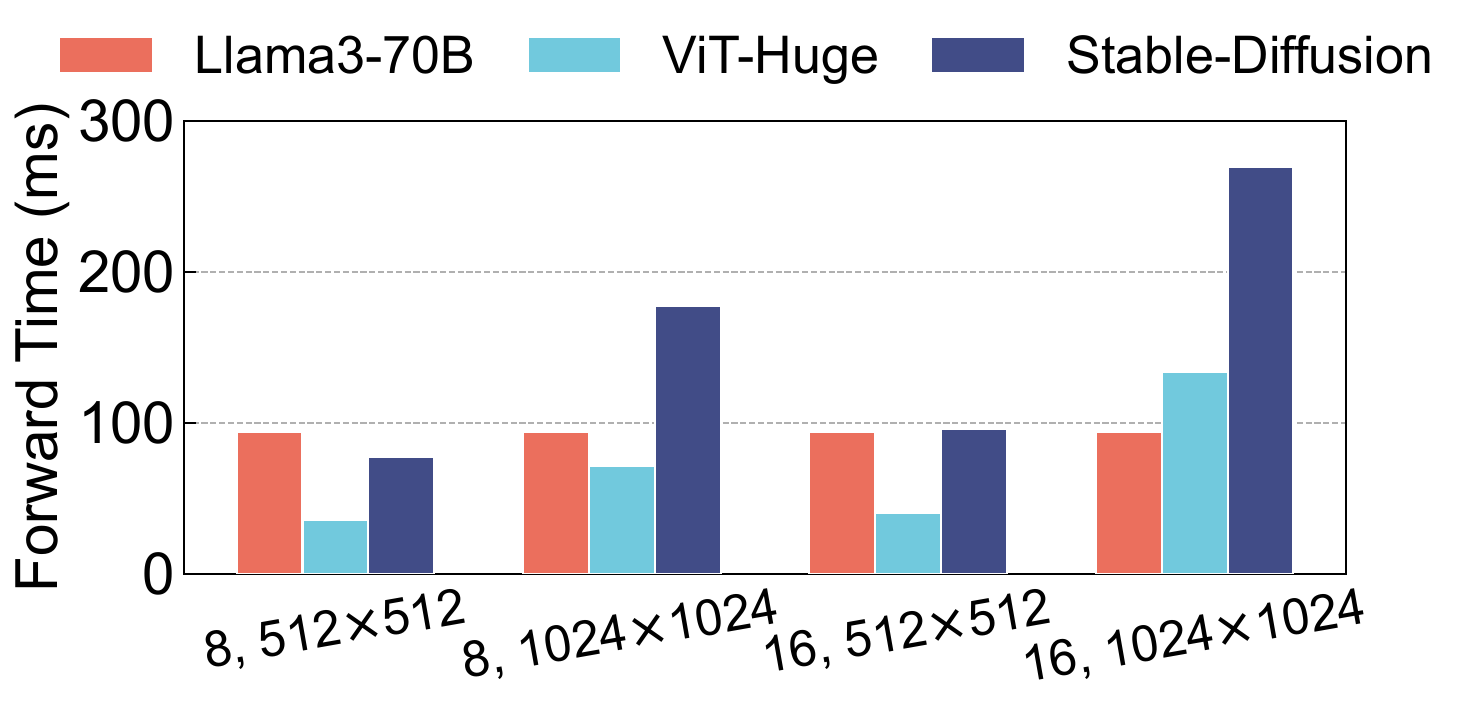}
    \vspace*{-0.15in}
    \caption{Forward time under different input configurations.}
    \vspace*{-0.0in}
    \label{fig:background:model_heter}
\end{figure}

\paraf{Characterization.}
Each module in multimodal LLMs has different computational demands due to
varying operators and inputs. For instance, ViT, as modality encoder, is
constructed with narrow transformer layers (i.e., small hidden size), whereas
the LLM backbone is built with wide transformer layers (i.e., large hidden
size). Meanwhile, Diffusion, as modality generator, utilizes a combination of
convolution and attention layers. The diversity in model architecture results in
distinct computation time for each module.
Figure~\ref{fig:background:model_heter} shows varying forward time under
different input configurations with Megatron-LM. We demonstrate one PP stage of
LLM backbone with PP size of 10 and TP size of 8. The first configuration
parameter is the number of images in the 8K input sequence, and the second is
the image resolution.
\revisionsigcomm{
The results reveal a key performance disparity: while the LLM backbone
maintains a consistent forward time, the modality encoder and generator exhibit significant variations.
}

\begin{figure}[t!]
    \centering
    \includegraphics[width=0.98\linewidth]{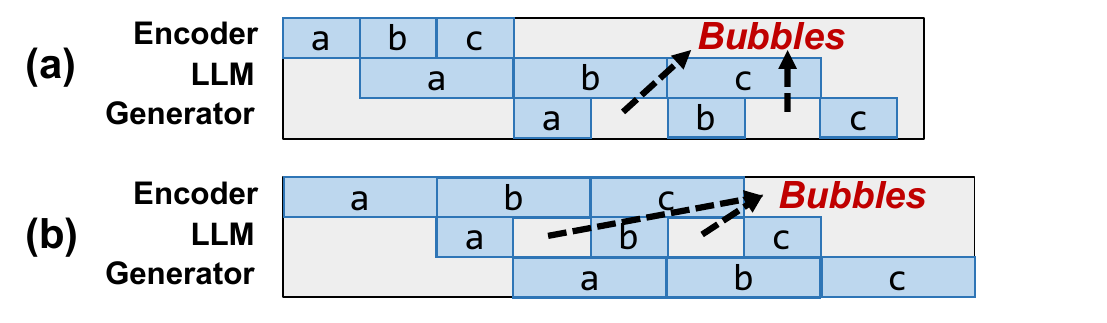}
    \vspace*{-0.15in}
    \caption{Two types of pipeline bubbles.}
    \vspace*{-0.1in}
    \label{fig:background:model_heter_pipe}
\end{figure}

\begin{figure*}[t!]
    \centering
    \includegraphics[width=\linewidth]{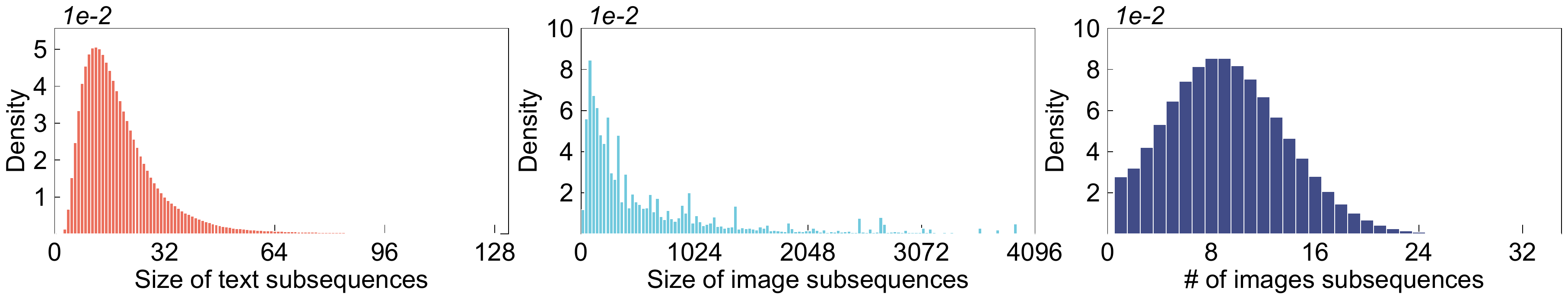}
    \small{\hspace*{0.13in}{(a) Distribution of text subsequence size.}\hspace*{\dimexpr\linewidth/26\relax}{(b) Distribution of image subsequence size.}\hspace*{\dimexpr\linewidth/30\relax}{(c) Distribution of image subsequence count.}}
    \vspace*{-0.06in}
    \caption{Data heterogeneity in multimodal LLM training.}
    \vspace*{-0.05in}
    \label{fig:background:data_heter}
\end{figure*}

\begin{figure}[t!]
    \centering
    \includegraphics[width=0.88\linewidth]{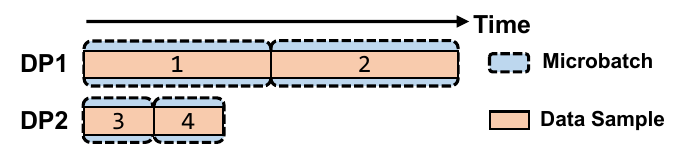}
    \vspace*{-0.08in}
    \caption{Intra-microbatch straggler (among DP groups).}
    \vspace*{-0.1in}
    \label{fig:background:data_heter_dp}
\end{figure}

\parabf{Pipeline imbalance.}
The computation imbalance between modules leads to two types of bubbles in
pipeline parallelism. The first arises in the modality encoder and generator
stages, as shown in
Figure~\ref{fig:background:model_heter_pipe}\textcolor{ACMDarkBlue}{(a)},
resulting from their inadequate utilization of allocated GPU resources. The
second type arises in the stages of the LLM backbone, as shown in
Figure~\ref{fig:background:model_heter_pipe}\textcolor{ACMDarkBlue}{(b)}.
This is because the intensive computation demands of the encoder and generator
increase their stage durations. Due to the pipeline dependency, the LLM stages
are forced to wait for the multimodal stage to complete, thereby creating
pipeline bubbles. The latter problem is particularly pronounced during
large-scale multimodal LLM training, where the bulk of GPU resources are
allocated to the LLM backbone. These pipeline bubbles, stemming from model
heterogeneity, reduce the MFU significantly during training.

\subsection{Data Heterogeneity}
\label{sec:background:data}

\paraf{Characterization.}
The training data samples of Multimodal LLMs often combine lightweight text with
heavyweight multimodal data. The latter significantly increases data
preprocessing time. For example, a typical training sample could include a
256-word text sequence and ten 1024$\times$1024 RGB images. The text is just
kilobytes, whereas the images are total of 120 megabytes. Preprocessing (e.g.,
decompression, resizing, and reordering) such samples can take several seconds,
and prolong the co-located training process. Additionally, each input sequence
consists of interleaved modality subsequences that exhibit highly skewed
distributions. Focusing on images and texts, we perform data characterization on
the LAION-400M dataset~\cite{schuhmann2021laion}. Each image (i.e., one image
subsequence) is segmented into 16$\times$16 patches, and each patch is converted
into one image token. The texts are tokenized through Llama tokenizer. The image
tokens are interleaved with text tokens to create an 8K-token training sequence.
As shown in
Figure~\ref{fig:background:data_heter}\textcolor{ACMDarkBlue}{(a)} and
Figure~\ref{fig:background:data_heter}\textcolor{ACMDarkBlue}{(b)}, the sizes
of text and image subsequences display highly skewed distributions. We further
analyze the count of modality subsequence per training sample using image as an
example. The count of image subsequences per training sample, shown in
Figure~\ref{fig:background:data_heter}\textcolor{ACMDarkBlue}{(c)}, also
shows a skewed distribution. Different sample sizes (i.e., modality tokens per
sample) lead to varying computation time in the modality encoder and generator.

\parabf{Stragglers.}
Data heterogeneity results in \emph{intra-microbatch} and
\emph{inter-microbatch} stragglers within the PP stages of the modality \emph{encoder}
and \emph{generator}. These stragglers exacerbate the computation imbalances and
further reduce GPU utilization. It is noted that all microbatches within the \emph{LLM}
have the same computation time since the sequence length is fixed. We do not
consider data heterogeneity between global batches, as each global batch
contains numerous randomly shuffled training samples (e.g., thousands with a
large DP size), which effectively smooths out data heterogeneity.

\begin{figure}[t!]
    \centering
    \includegraphics[width=0.94\linewidth]{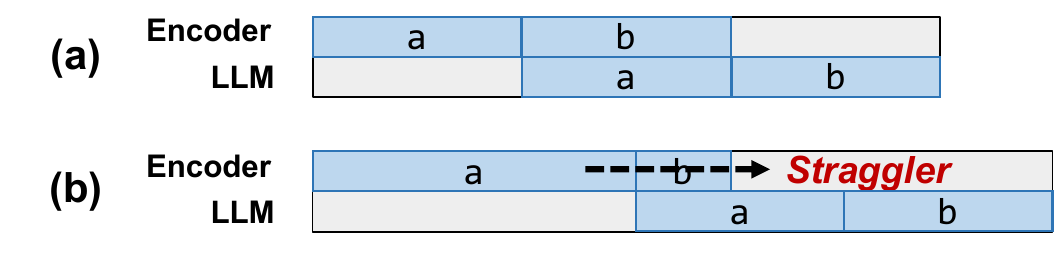}
    \vspace*{-0.12in}
    \caption{Inter-microbatch straggler.}
    \vspace*{-0.1in}
    \label{fig:background:data_heter_pipe}
\end{figure}

\parait{\underline{Intra-microbatch straggler}}
arises since different DP groups handle variably-sized training samples. Illustrated in Figure~\ref{fig:background:data_heter_dp},
the first DP group ($\mathit{DP1}$) processes two large training samples within two microbatches. In contrast,
the second DP group ($\mathit{DP2}$) processes two smaller samples in the same microbatches, completing them more quickly.
Consequently, $\mathit{DP1}$ lags behind $\mathit{DP2}$ and becomes the straggler, which delays the overall training process.

\parait{\underline{Inter-microbatch straggler}}
emerges from pipeline imbalances between microbatches.
As depicted in Figure~\ref{fig:background:data_heter_pipe}, the first pipeline stage is the modality encoder followed by one LLM backbone stage.
Figure~\ref{fig:background:data_heter_pipe}\textcolor{ACMDarkBlue}{(a)}
illustrates the pipeline without data heterogeneity, where the modality encoder processes each microbatch with the same amount of time.
In contrast, Figure~\ref{fig:background:data_heter_pipe}\textcolor{ACMDarkBlue}{(b)}
depicts the pipeline with data heterogeneity, where the forward time of the modality encoder
varies largely across microbatches. The straggler (i.e., the microbatch \texttt{a}) significantly
delays the training process of the subsequent PP stages, leading to a large pipeline bubble.
\section{\sysname Overview}
\label{sec:overview}

\begin{figure}[t]
    \centering
    \includegraphics[width=0.86\linewidth]{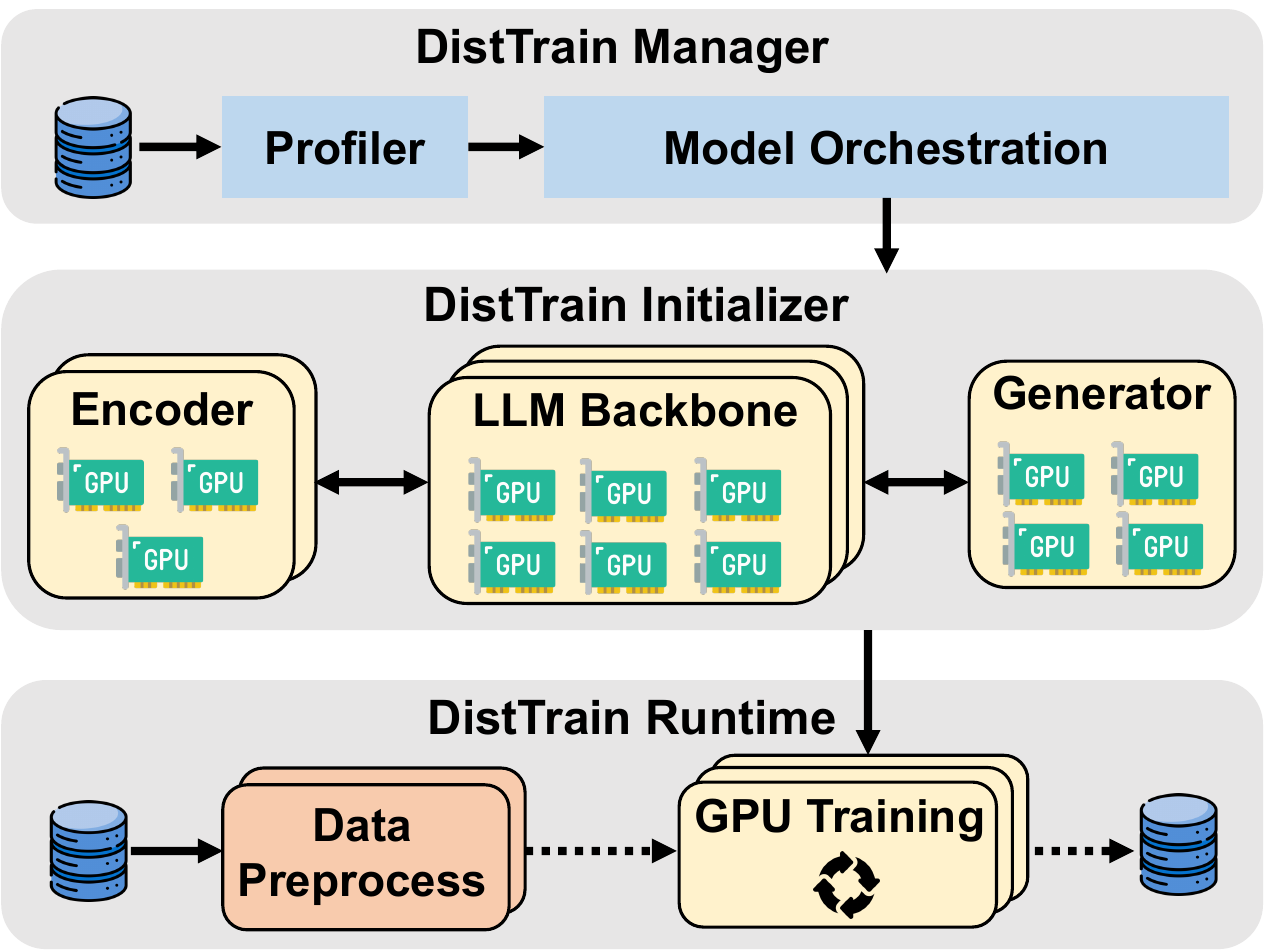}
    \vspace{-0.1in}
    \caption{\sysname overview.}
    \vspace{-0.1in}
    \label{fig:overview:arch}
\end{figure}

We propose \sysname, a disaggregated training system for multimodal LLMs.
\sysname addresses model heterogeneity by disaggregated model orchestration
(\S\ref{sec:design:resource}) and handles data heterogeneity by disaggregated
data preprocessing (\S\ref{sec:design:reorder}). Figure~\ref{fig:overview:arch}
shows an overview of \sysname.

\parabf{\sysname manager.}
Before training, \sysname employs a training manager to determine the resource
allocation and parallelism strategy for each module in multimodal LLMs. The
training manager first gathers the model architecture and training configuration
(e.g., global batch size) from the user and samples a subset of training data to
analyze the data distribution. Utilizing the information, it runs a series of
benchmarking training trials and constructs a performance profiler with linear
interpolation to estimate each module's computation and communication time.
Based on the profiling results, the training manager decides the optimal
resource allocation and parallelism strategy with \emph{disaggregated model
orchestration} for one specific training task, as detailed in
\S\ref{sec:design:resource}.

\parabf{\sysname initializer.}
\sysname then initializes the disaggregated model orchestration for
modality encoder, LLM backbone, and modality generator, respectively.
\sysname allocates different numbers of GPUs and determines the parallelism strategy for each module.
Each module then establishes the specific communication group.
It then loads the model checkpoint from a distributed file system
and shards the model parameters.
Finally, \sysname conducts several communication trials to warm up the system and test connectivity.

\parabf{\sysname runtime.}
At runtime, the dedicated CPU nodes (i.e., disaggregated data preprocessing nodes)
retrieve training data samples from the distributed file system for preprocessing.
It performs \emph{data reordering} to reorder the training samples within one global batch
while preserving the synchronous training semantics~\cite{alpa-tu}.
The reordering mitigates both intra-microbatch and inter-microbatch stragglers caused by data heterogeneity,
as detailed in \S\ref{sec:design:reorder}. In each iteration,
the GPU training nodes receive the preprocessed data asynchronously from the CPU nodes.
The data then undergoes sequentially through the modality encoder, LLM backbone, and modality generator in the training pipeline.
Finally, the GPU training nodes synchronize the gradients and model parameters through collective communication.
\revisionsigcomm{
\sysname employs ZERO-1 optimization~\cite{rajbhandari2020zero} and mixed precision training~\cite{micikevicius2017mixed}
to reduce the memory footprint and accelerate the training of LLM backbone.
Additionally, \sysname adopts a dedicated process to periodically and asynchronously save
model checkpoints to the distributed file system for fault tolerance.
}

\section{Addressing Model Heterogeneity}
\label{sec:design:resource}

To address model heterogeneity, we first introduce \emph{disaggregated model
orchestration} to adaptively orchestrate the resource and parallelism
configurations of the three modules in multimodal LLM training. We then
formulate the problem to minimize the training time per iteration. Last, we
present the adaptive model orchestration algorithm to find optimal resource and
parallelism configurations.

\subsection{Disaggregated Model Orchestration}
\label{sec:design:resource:dis}

\begin{figure}[t!]
    \centering
    \includegraphics[width=0.98\linewidth]{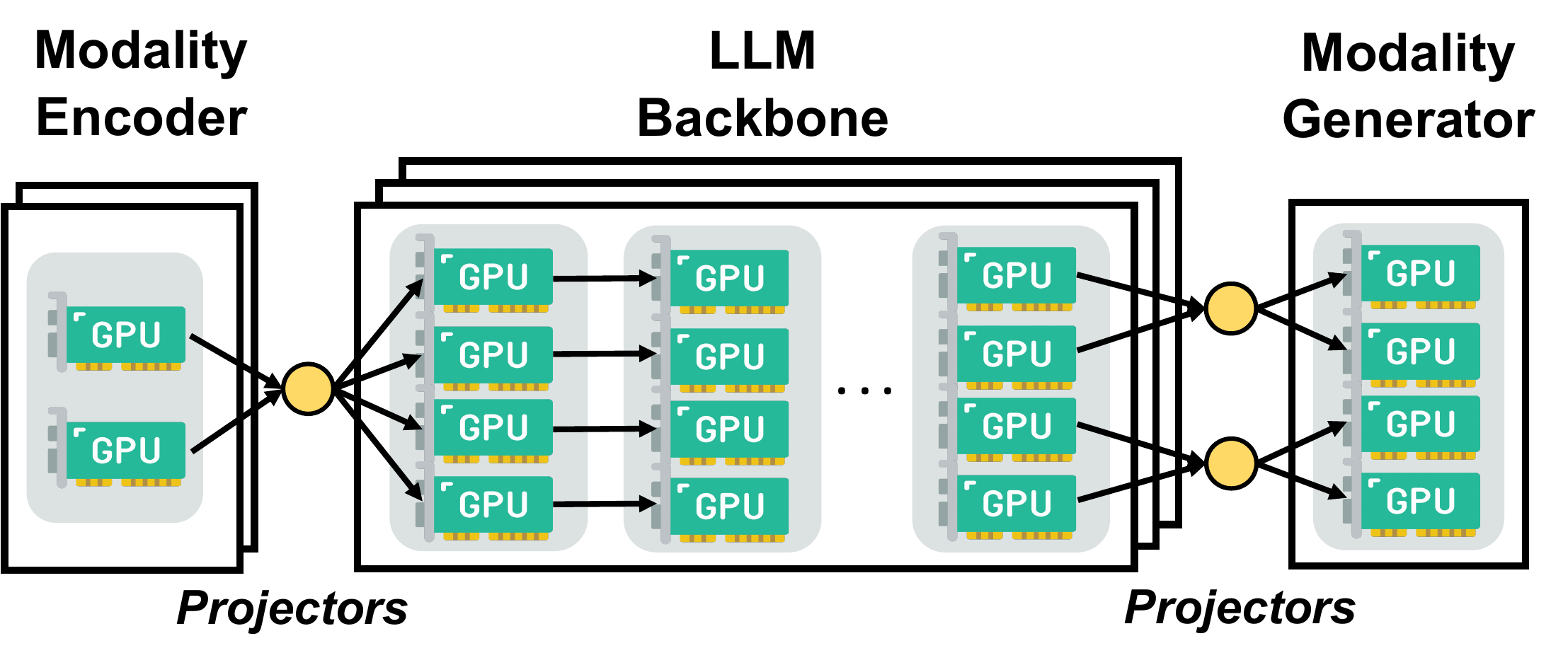}
    \vspace*{-0.2in}
    \caption{Disaggregated model orchestration in \sysname.}
    \vspace*{-0.15in}
    \label{fig:design:dis_training}
\end{figure}

Figure~\ref{fig:design:dis_training} illustrates the disaggregated model
orchestration. Different from the monolithic model orchestration in Megatron-LM
(i.e., Figure~\ref{fig:background:megatron_training}), \sysname is able to
adaptively adjust the resource allocation and parallelism strategy. For
instance, \sysname can allocate 4 GPUs (DP=2 and TP=2) to the modality encoder,
12 GPUs (DP=3 and TP=4) to the LLM backbone per PP stage, and 4 GPUs (DP=1 and
TP=4) to the modality generator. Additionally, the projector layers are
co-located with either the modality encoder or generator, with their number of
replicas adapting as needed. We implement disaggregated model orchestration
through a dedicated module, i.e., \emph{parallelism unit}.

\parabf{Parallelism unit.}
At training initialization, we need to establish the communication group
according to the resource allocation and parallelism strategy. \sysname
introduces a module, parallelism unit, composed of one or more PP stages. Each
unit can adopt its own DP and TP strategies and form a specific communication
group. Inter-unit connections are facilitated by a \emph{communication broker},
which bridges PP communication across parallelism units. Users are only required
to specify the DP and TP configurations for each parallelism unit, and \sysname
automatically sets up the communication group and communication broker. \sysname
treats the modality encoder, LLM backbone, and modality generator as three
individual parallelism units.
\revisionsigcomm{
For advanced workloads, \sysname extends the capabilities of these units.
To handle long sequences, it integrates sequence parallelism (SP) within
the LLM backbone unit and automatically splits sequences within its SP group.
Furthermore, \sysname supports expert parallelism (EP) for the LLM backbone.
Since EP and TP both perform parallel computation and communication within one layer,
our subsequent formulation involving TP remains valid when TP is replaced with EP.
}
The detailed implementation of parallelism unit is
described in \S\ref{sec:implementation}.

\subsection{Problem Formulation}
\label{sec:design:resource:form}

With disaggregated model orchestration, we are able to adaptively orchestrate the three modules.
The problem now lies in determining the optimal resource allocation and parallelism strategy to minimize the training time per iteration.
Exhaust search is infeasible due to the combinatorial search space, particularly for large clusters.
One strawman solution is to allocate the resources proportional to the model flops of each module.
However, this method falls short as it overlooks complex patterns in parallelism training.
\revisionsigcomm{
Different DP and TP sizes in modality encoder could lead to different computation times while the model flops are the same.
}
Before diving into our solution, we first formulate the optimization problem.

\parabf{LLM backbone.}
We begin by formulating the LLM backbone, focusing on the forward pass as the backward pass mirrors it.
\revisionsigcomm{
In production LLM training, the microbatch size is set to a small number to prevent GPU memory overflow.
\sysname sets the microbatch size to a predefined constant, $M$.
}
Assume the global batch size for one iteration as $BS$ and the TP size of the LLM backbone as $TP_{lm}$.
Let the PP and DP size of the LLM backbone be $PP_{lm}$ and $DP_{lm}$.
\revisionsigcomm{
The LLM is divided into equal $PP_{lm}$ PP stages, where each PP stage consists of some homogeneous transformer layers.
}
The number of GPUs allocated to the LLM backbone is $y=TP_{lm} \times DP_{lm} \times PP_{lm}$.
\revisionsigcomm{
Let the forward time (including communication time) of the entire LLM for one sample be $C_{lm}(TP_{lm})$, where $C_{lm}$ represents the forward time function.
Therefore, the forward time of one PP stage for one microbatch with $M$ samples is $T_{lm}=\frac{C_{lm}(TP_{lm}) \times M}{PP_{lm}}$ since
PP partitions LLM backbone's homogeneous layers sequentially.
Besides, the number of microbatches per iteration is $\frac{BS}{DP_{lm} \times M}$.
}

\parabf{Modality encoder and generator.}
In \sysname, the modality encoder is regarded as a parallelism unit with PP size $PP_{me}$.
Let the TP size be $TP_{me}$ and the DP size be $DP_{me}$.
The number of GPUs allocated to the modality encoder is $x=TP_{me} \times DP_{me} \times PP_{me}$.
\revisionsigcomm{
The microbatch size is $\frac{DP_{lm} \times M}{DP_{me}}$ which is determined by the LLM backbone.
}
Let the forward time of the entire modality encoder for one training sample be $C_{me}(TP_{me})$.
\revisionsigcomm{
The forward time of one PP stage for one microbatch in the modality encoder is $T_{me}=\frac{DP_{lm} \times M}{DP_{me}} \times \frac{C_{me}(TP_{me})}{PP_{me}} = \frac{DP_{lm} \times TP_{me} \times M}{x} \times C_{me}(TP_{me})$.
Similarly, the forward time of one PP stage in the modality generator is $T_{mg}= \frac{DP_{lm} \times TP_{mg} \times M}{z} \times C_{mg}(TP_{mg})$,
}
where $z$ is the number of GPUs allocated to modality generator.

\begin{figure}[t!]
    \centering
    \includegraphics[width=0.9\linewidth]{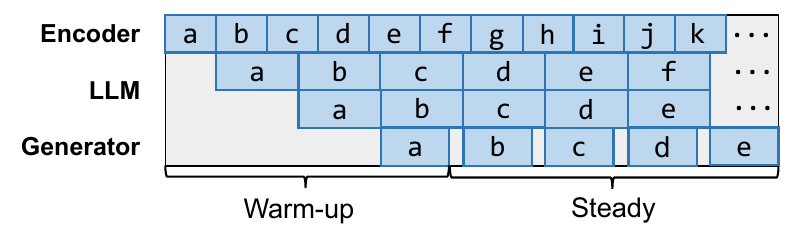}
    \vspace*{-0.13in}
    \caption{Multimodal LLM training pipeline.}
    \vspace*{-0.18in}
    \label{fig:design:pipeline_obj}
\end{figure}

\parabf{Objective function.}
Based on the preceding analysis, we next define the objective function for the optimization problem, i.e.,
the training time of one iteration.
Figure~\ref{fig:design:pipeline_obj}
shows the pipeline of forward pass in multimodal LLM training. The LLM backbone comprises two PP stages, whereas the modality encoder and generator each consist of one PP stage.
The forward pass is categorized into two phases: warm-up phase and steady phase. 
The warm-up phase spans from the initiation to the completion of the first microbatch to populate the pipeline.
The duration of this phase is calculated as $T_{warmup}=T_{lm} \times PP_{lm}  + T_{me} \times PP_{me} + T_{mg} \times PP_{mg}$,
which is formulated as follows.
\revisionsigcomm{
\begin{align}\label{formula:design:f1}
    \nonumber
    T_{warmup} &= M \times C_{lm}(TP_{lm}) + \frac{DP_{lm} \times M}{DP_{me}} \times C_{me}(TP_{me}) \\
    &+ \frac{DP_{lm} \times M}{DP_{mg}} \times C_{mg}(TP_{mg})
\end{align}
}
The duration of steady phase is dominated by the maximal computation time among PP stages,
which is calculated as $T_{steady}=\max(T_{lm},T_{me}, T_{mg}) \times (\frac{BS}{DP_{lm} \times M}-1)$,
where $\frac{BS}{DP_{lm} \times M}$ is the number of microbatches per iteration.
It is formulated as:
\revisionsigcomm{
\begin{align}\label{formula:design:f2}
    \nonumber
    &T_{steady} \\
    &= max \left\{
        \begin{array}{lll}
            \frac{DP_{lm} \times TP_{lm} \times M}{y} \times C_{lm}(TP_{lm}), \\
            \frac{DP_{lm} \times TP_{me} \times M}{x} \times C_{me}(TP_{me}), \\
            \frac{DP_{lm} \times TP_{mg} \times M}{z} \times C_{mg}(TP_{mg})   
        \end{array}
    \right\} \times (\frac{BS}{DP_{lm} \times M}-1)
\end{align}
}
Therefore, the objective function is to minimize $T_{warmup} + T_{steady}$.
\revisionsigcomm{
Considering backward pass with the similar computation pattern, the objective function is
similar to that of the forward pass.} Adjustments are made by changing
$C_{lm}, C_{me}$, and $C_{mg}$ from forward time functions to 
the sum functions of forward and backward time.
This formulation holds for GPipe~\cite{huang2019gpipe} and 1F1B~\cite{fan2021dapple}. We will retrofit the formulation to adapt to VPP~\cite{narayanan2021efficient} later.
TP communication is incorporated into the functions $C_{lm}, C_{me}$, and $C_{mg}$, which are calibrated through interpolation from actual trials.
The communication time of DP and PP is modeled as the communication volume divided by the bandwidth.

\parabf{Constraints.}
Besides the objective function, we must consider constraints to ensure training feasibility.
The first constraint is the resource constraint.
The number of GPUs allocated to each module should be
$x + y + z \leq N$
where $N$ is the total number of GPUs in the cluster
and $x,y,z$ are the number of GPUs allocated to each module in multimodal LLM.
\revisionsigcomm{
The second constraint involves GPU memory.
}
We first consider LLM backbone.
Memory allocation involves four parts: model parameters, gradients, optimizer states, and activation states.
The memory of model parameters and gradients on one GPU is calculated as: $\frac{P}{PP_{lm}\times TP_{lm}} = \frac{DP_{lm} \times P}{y}$,
where $P$ denotes the total memory for the LLM parameters and gradients.
The memory for optimizer states on one GPU (with ZeRO-1 optimization~\cite{rajbhandari2020zero}) is: $\frac{S}{y}$, where $S$ denotes the total memory for the optimizer states.
ZeRO-1 partitions the optimizer states across DP groups.
The peak memory for activation states on one GPU is: $\frac{DP_{lm} \times L \times PP_{lm}}{y}$, with $L$ representing the memory needed for one
microbatch \revisionsigcomm{(with microbatch size $M$)} of activation states across the entire LLM. In 1F1B, the first PP stage requires storage for
$PP_{lm}$ microbatches of activation states.
\revisionsigcomm{
We do not use GPipe in \sysname since it consumes more memory
without offering better training efficiency compared to 1F1B.
}
The memory constraint ensures the sum of the four memory parts on one GPU does not exceed GPU capacity.
As for the modality encoder and generator, the formulation is similar.

\subsection{Adaptive Model Orchestration}
\label{sec:design:resource:sloving}
The problem is non-convex, with $x, y, z$, and the DP, TP sizes as positive variables.
Solving this with an exhaust search is impractical due to the large search space, particularly in large clusters.
Designing an efficient algorithm that \emph{adaptively} identifies the optimal resource allocation and
parallelism strategy based on the training task poses a hard problem.

\begin{figure}[t!]
    \centering
    \includegraphics[width=0.76\linewidth]{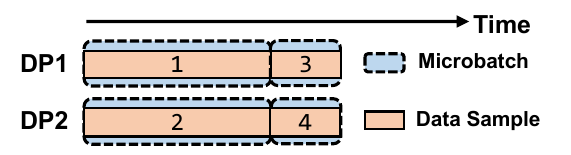}
    \vspace*{-0.15in}
    \caption{Intra-microbatch reordering.}
    \vspace*{-0.1in}
    \label{fig:design:intra_reorder}
\end{figure}

\parabf{Convex optimization.}
Our key insight is to decompose the non-convex optimization problem into a series of simplified convex problems
with variables $x,y,z$ (i.e., resource allocation).
We confine the TP size to $[1, 2, 4, 8]$ on an NVIDIA GPU node with 8 GPUs and adjust the DP size as a factor of $BS$ to balance the computation across DP groups.
\revisionsigcomm{
If expert parallelism (EP) is used instead of TP, we confine EP size to factors of cluster size.
}
The PP size of the LLM backbone is calculated as $\frac{y}{DP_{lm} \times TP_{lm}}$.
The set of possible parallelism strategies is a manageable and finite set, i.e., the Cartesian product of TP and DP size.
This allows us to enumerate all feasible TP and DP sizes, and transform the original optimization problem into a set of simplified problems.
In the simplified problem, the objective function is maximal and additional of the functions: $\frac{1}{x}$, $\frac{1}{y}$ and $\frac{1}{z}$,
where $x,y,z$ are positive variables. Therefore, the objective function is convex.
Similarly, the constraint functions are also convex.
As a result, the simplified optimization problem is convex and can be efficiently solved to optimality by off-the-shelf solvers~\cite{grant2014cvx, virtanen2020scipy}.
The algorithm \emph{adaptively} determines the optimal resource allocation and parallelism strategy.

Virtual pipeline parallelism~\cite{narayanan2021efficient} (i.e., interleaved 1F1B) reduces the warm-up time
by dividing model into finer-grained virtual PP (VPP) stages. Each PP stage contains VPP-size VPP stages.
In the warm-up phase, each PP stage launches the computation of one VPP stage, and the warm-up time is divided by VPP-size.
To align our formulation with VPP, we proportionally reduce the warm-up time based on VPP-size.

\section{Addressing Data Heterogeneity}
\label{sec:design:reorder}

To address data heterogeneity, we first introduce \emph{disaggregated data
preprocessing} that separates data preprocessing from training. This
eliminates the interference between data processing and training
and enables negligible overhead of data processing. We then
introduce \emph{disaggregated data reordering} to address
data heterogeneity. We use intra-microbatch reordering to eliminate stragglers
across DP groups, and inter-microbatch reordering to minimize pipeline bubbles
and hide the training stragglers into the training pipeline as much as possible.

\subsection{Disaggregated Data Preprocessing}
\label{sec:design:reorder:dis}

As discussed in \S\ref{sec:background:data}, preprocessing high-volume multimodal data (e.g., high resolution images)
incurs substantial overhead.
\sysname disaggregates data preprocessing from training with a producer-consumer model.
The producer, operating on dedicated CPU nodes, fetches data from the distributed file system and preprocesses
training data asynchronously. 
The consumer, i.e., the main training process on dedicated GPU nodes, receives the preprocessed
data for training.
The producer and consumer communicate through RPC calls, and use RDMA for lower latency if available.

The disaggregation enables flexible and elastic data preprocessing and provides opportunities for data reordering without incurring additional overhead.
Leveraging this opportunity, we incorporate \emph{data reordering} into data preprocessing to mitigate training stragglers
caused by data heterogeneity. The reordering includes two levels: intra-microbatch and inter-microbatch.
We emphasize that disaggregated data preprocessing ensures that the complex reordering does not interfere with the GPU training or impose extra overhead.

\subsection{Intra-microbatch Reordering}
\label{sec:design:reorder:intra}
\paraf{Insight.}
To address intra-microbatch stragglers,
we first identify the straggler by pinpointing the DP group with the largest training samples.
As shown in Figure~\ref{fig:background:data_heter_dp}, the first DP group becomes a straggler as it contains
the two largest training samples. Therefore, we reorder the
training samples within the global batch by size. Specifically, as depicted in Figure~\ref{fig:design:intra_reorder},
we reorder the training samples into the sequence $[1,3,2,4]$, which distributes
the computation more evenly.

\begin{figure*}[t!]
    \centering
    \includegraphics[width=1\linewidth]{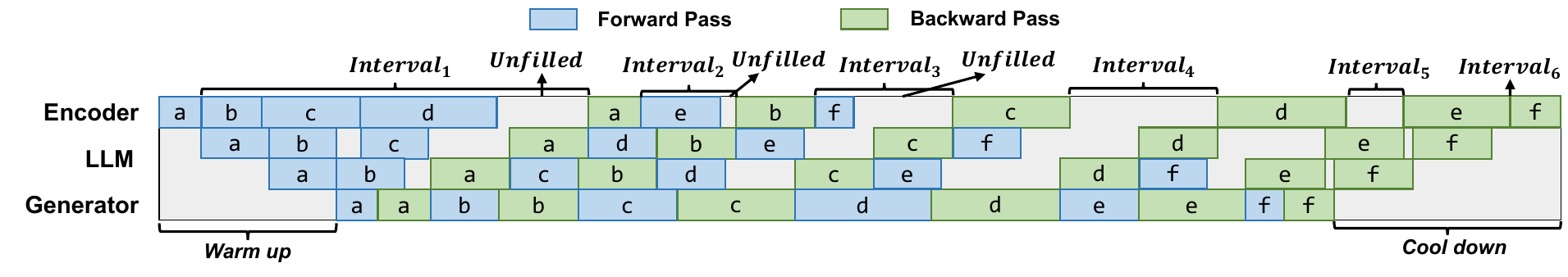}
    \vspace*{-0.35in}
    \caption{1F1B pipeline scheme.}
    \vspace*{-0.0in}
    \label{fig:design:inter_reorder}
\end{figure*}

\parabf{Intra-microbatch Reordering.}
\revisionsigcomm{
Leveraging this insight, we propose intra-microbatch reordering to balance the computational load within a global batch. 
Formally, the goal is to minimize the maximum computation time across data parallelism (DP) groups.
This problem maps to the multiway number partitioning problem~\cite{korf2009multi}, which is NP-hard.
Given the problem's complexity and the large batch sizes common in production,
finding an optimal solution is computationally infeasible.
We therefore employ a lightweight, greedy partitioning algorithm that yields an approximation ratio of
less than 4/3~\cite{barman2020approximation}.
Importantly, this reordering only occurs within the global batch, which only
affects the sequence of gradient accumulation. Since gradient accumulation (i.e., summation)
is a commutative operation, the intra-microbatch reordering preserves
the training's convergence semantics.
}

\revisionsigcomm{
The detailed algorithm is summarized in Algorithm~\ref{alg:design:intra_reorder}.
The function $\mathit{INTRAREORDER}$ receives the $n$ original training samples and DP size $m$.
This algorithm first sorts the training samples in descending order by the sample size (line~3).
Then, it loops over the training samples and assigns the sample to the DP group with the current lowest computational load (line~6-8).
It then returns the reordered samples (line~9-11).
The algorithm has a time complexity of $O(n \log n + m \times n)$.
}

\begin{algorithm}[t]
    \caption{\revisionsigcomm{Intra-batch reordering.}}
    \label{alg:design:intra_reorder}
    \begin{normalsize}
    \begin{algorithmic}[1]
        \Function {INTRAREORDER}{$\{d_1, ..., d_n\}$, $m$}
            \State $\mathit{sorted\_samples} \gets \{d_1, ..., d_n\}$, $\mathit{ret\_samples} \gets \emptyset$, $\mathit{Groups} \gets \emptyset$
            \State Sort $\mathit{sorted\_samples}$ in descending order based on $\mathit{d_i.size}$
            \For{$i=1 \rightarrow m$}
                \State $\mathit{Group_i} \gets \emptyset$, $\mathit{Groups.append(Group_i)}$
            \EndFor
            \For{$i=1 \rightarrow n$}
                \State $\mathit{min\_index} \gets argmin_{j} \sum_{d \in Group_j} d.size$
                \State $\mathit{Groups[min\_index].append(sorted\_samples[i])}$
            \EndFor
            \For{$i=1 \rightarrow m$}
                \State $\mathit{ret\_samples.extend(Groups[i])}$
            \EndFor
            \State \Return $\mathit{ret\_samples}$
        \EndFunction
    \end{algorithmic}
    \end{normalsize}
\end{algorithm}

\subsection{Inter-microbatch Reordering}
\label{sec:design:reorder:inter}
As we discussed in \S\ref{sec:background:data}, data heterogeneity leads to varied computation times across
microbatches within the modality encoder and generator.
The straggler microbatch prolongs the training by creating large pipeline bubbles.
In the context of 1F1B pipeline scheme, the overall iteration time is primarily decided by pipeline
bubbles and the computation time at the first PP stage of the modality encoder, as illustrated in Figure~\ref{fig:design:inter_reorder}.
Let the PP size be $p$ and the number of microbatches be $l$ ($p=4$ and $l=6$ in Figure~\ref{fig:design:inter_reorder}).
Typically, $l$ is larger than $p$ to reduce the proportion of time spent in the warm-up and cool-down phases.
To help describe our solution, we first introduce 
a concept called pipeline \emph{intervals} at the first PP stage. As shown in Figure~\ref{fig:design:inter_reorder},
these intervals are typically filled with the forward pass, except for the last $p-1$ intervals (i.e., $\mathit{interval_{4}}$, $\mathit{interval_{5}}$, and $\mathit{interval_{6}}$).
Straggler microbatches in either the encoder or generator prolong these intervals or
increase the unfilled area (i.e., bubble).
Designing an efficient runtime algorithm to minimize pipeline bubbles presents a significant challenge.

\parabf{Insights.}
We leverage two insights to solve this problem. The first insight involves minimizing the volume of intervals that are not filled.
As shown in Figure~\ref{fig:design:inter_reorder}, the last
$p-1$ intervals (i.e., $\mathit{interval_{4}}$ to $\mathit{interval_{6}}$) remain unfilled.
These intervals become the pipeline bubbles and increase the training iteration.
We observe a positive correlation between the volume of $interval_{i}$ and the size of the $i_{th}$ microbatch.
The size refers to the computation time of the microbatch in modality encoder and generator.
For instance, $interval_{4}$ is significantly larger than $interval_{5}$ and $interval_{6}$
since the $4_{th}$ microbatch (i.e., \texttt{d}) is the largest.
By strategically reordering the training samples to position the smallest $p-1$ microbatches at the end,
we are able to reduce these unfilled intervals' volume.

The second insight involves minimizing the unfilled area of the left intervals,
which hides the training stragglers into the training pipeline as much as possible.
The left intervals (i.e., $\mathit{interval_{1}}$ to $\mathit{interval_{3}}$) are filled with the forward pass.
As shown in Figure~\ref{fig:design:inter_reorder}, the first interval is filled with by the $2_{nd}$ to $p_{th}$ forward passes.
For subsequent intervals, $\mathit{interval_{i}}$ is filled by the $(i+p-1)_{th}$ forward pass.
However, the forward pass may not perfectly match the interval volumes, leading to unfilled areas.
By evaluating the volume of $\mathit{interval_{i}}$, we place the microbatches, whose forward time most closely matches this volume,
at the corresponding position, to minimize the unfilled area.

\begin{algorithm}[t]
    \caption{Inter-batch reordering.}
    \label{alg:design:inter_reorder}
    \begin{normalsize}
    \begin{algorithmic}[1]
        \Function {INTERREORDER}{$\{m_1, ..., m_l\}$, $p$}
            \State $\mathit{ret\_mb} \gets \emptyset$, $\mathit{mb} \gets \{m_1, ..., m_l\}$
            \State $\mathit{ret\_mb.append(MIN(mb))}$, $\mathit{mb.remove(MIN(mb))}$
            \State $\mathit{rear\_mb \gets SELECTMIN(mb, p-1)}$, $\mathit{mb.remove(rear\_mb)}$
            \For{$i=1 \rightarrow l-p$}
                \State $\mathit{interval_i} \gets \mathit{GETINTERVAL(ret\_mb, i)}$
                \If {$i==1$}
                    \State $\mathit{cur\_mb} \gets \mathit{SELECTCLOSEST(mb, p-1, interval_i)}$
                \Else
                    \State $\mathit{cur\_mb} \gets \mathit{SELECTCLOSEST(mb, 1, interval_i)}$
                \EndIf
                \State $\mathit{ret\_mb.extend(cur\_mb)}$, $\mathit{mb.remove(cur\_mb)}$
            \EndFor
            \State $\mathit{ret\_mb.extend(rear\_mb)}$
            \State \Return $\mathit{ret\_mb}$
        \EndFunction
    \end{algorithmic}
    \end{normalsize}
\end{algorithm}

\parabf{Inter-microbatch Reordering.}
\revisionsigcomm{
Based on the preceding insights, we propose a runtime algorithm for inter-microbatch
reordering designed to minimize pipeline bubbles. This algorithm is specifically developed for
the 1F1B pipeline schedule. We will retrofit the algorithm
to VPP (i.e., interleaved 1F1B) later. Algorithm~\ref{alg:design:inter_reorder} summarizes the pseudo code.
The $\mathit{INTERREORDER}$ function accepts the initial microbatch sequence and the PP size $p$.
The reordering process begins by scheduling the smallest microbatch first to ensure all pipeline stages are activated promptly (line~3).
Subsequently, it reserves the $p-1$ smallest remaining microbatches for the end of
the sequence to minimize unfilled intervals (lines~4 and line~12).
The main loop (lines~5-11) then iterates through the remaining microbatches to fill the pipeline bubbles.
In each loop iteration, it computes the interval size via the $\mathit{GETINTERVAL}$ function and
applies a heuristic to select microbatches that best fit this size.
For the first interval, it greedily selects $p-1$ microbatches whose aggregate forward time
most closely matches the interval size; for all subsequent bubbles,
it selects the single best-fitting microbatch.
This loop ensures maximal filling of the remaining intervals, which minimizes pipeline bubbles.
Crucially, this reordering only alters the data sequence within the local batch of a single DP rank per
training iteration. Therefore, analogous to intra-microbatch reordering,
it only affects the sequence of gradient accumulation and consequently preserves the training's convergence semantics.
}

The functions $\mathit{SELECTMIN}$ and $\mathit{SELECTCLOSEST}$ operate with a time complexity of $O(l)$.
The function $\mathit{GETINTERVAL}$ calculates interval size using the current order $\mathit{ret\_mb}$.
This calculation is facilitated by a dynamic programming algorithm that utilizes a recursive formula derived from pipeline dependencies.
Specifically, the start time of each microbatch depends on two factors: the completion of the preceding microbatch in the same pipeline stage
and the availability of input data from the upstream microbatch. Consequently, the end time of each microbatch is
determined by the maximum of these two dependencies plus its own computation time.
This dynamic programming algorithm exhibits a complexity of $O(p)$ per function invocation.
The algorithm has a time complexity of $O(l \times (l+p))$.

Virtual pipeline parallelism (i.e., interleaved 1F1B) also follows the one forward and one backward pipeline scheme
to reduce the memory footprint. The fundamental insights of our algorithm apply to any 1F1B-based pipeline, including VPP.
We adapt the algorithm by computing multiple (i.e., VPP size) intervals and filling them
with the corresponding number of forward passes from a single microbatch.
\section{System Implementation}
\label{sec:implementation}

We implement \sysname with 6.3K lines of code in C++ and Python, and integrate it with Megatron-LM~\cite{shoeybi2019megatron}.
\sysname handles failures by automatically
recovering the training from the latest model checkpoint.
To mitigate Tensor Parallelism (TP) overhead, we developed StepCCL, an in-house
collective communication library. By leveraging the DMA engine for data transfers,
StepCCL overlaps communication with computation, which frees the core Streaming Multiprocessors (SMs)
to execute computation kernels (e.g., GEMM) without performance interference. 
A detailed description and evaluation are in \S\ref{appendix:ccl}.

\parabf{\sysname manager and initializer.}
\sysname's training manager, running on a dedicated CPU node,
formulates the disaggregated model orchestration problem using Disciplined Convex Programming~\cite{grant2006disciplined}.
It employs the CVX solver~\cite{cvxpy} to solve this problem within a second.
The manager records the optimal resource allocation and parallelism strategy to a configuration file,
which the Kubernetes controller uses to launch the training task.
As for the initializer,
\sysname uses PyTorch Distributed~\cite{torchdist} library to initialize the communication groups.
The initialization of each module in a multimodal LLM is implemented through parallelism units.


\parabf{Parallelism unit.}
\revisionsigcomm{
As discussed in~\S\ref{sec:design:resource:dis}, disaggregated model orchestration
is implemented via a dedicated module: \emph{parallelism unit}. 
During distributed training initialization, \sysname first establishes
communication groups within a parallelism unit. Each GPU process possesses a global and a local rank
within its unit, facilitating distributed initialization. Subsequently, \sysname
initializes a communication broker to establish PP communication between adjacent parallelism units.
All inter-unit communication traffic is routed via the communication broker. 
The communication broker is implemented by modifying Megatron-LM's batched
\texttt{send}/\texttt{receive} operations into discrete operations.
It enables flexible communication between multiple upstream and downstream GPU processes
by concentrating and scattering data as needed, while preserving data order.
The communication broker employs a decentralized design, residing on the GPU of either
the last PP stage in an upstream unit or the first PP stage in a downstream unit.
To maximize communication bandwidth, the number of brokers between two units is determined
by the greatest common divisor of their respective DP sizes. Consequently,
the total inter-unit bandwidth scales effectively with the training workload,
preventing the communication broker from becoming a training bottleneck.
Moreover, Megatron-LM's reliance on synchronous communication compels upstream
stages to pause until downstream stages fully receive data, introducing
unnecessary pipeline dependencies. To mitigate this, we implement asynchronous
\texttt{send} operations, eliminating these superfluous dependencies,
and redesign the communication topology to prevent potential deadlocks.
}
\section{Evaluation}
\label{sec:evaluation}

In this section, we first use large-scale experiments to evaluate the overall
performance improvements of \sysname over Megatron-LM.
Next, we use an \revisionsigcomm{ablation study} to deep dive into \sysname and show the effectiveness
of each technique.
Finally, we provide a case study to further evaluate \sysname.

\parabf{Setup.}
Our experiments are conducted on a production cluster, with each
node equipped with 8 NVIDIA Ampere GPUs.
GPUs within one node are
interconnected by 300GB/s (bidirectional) NVLink, while nodes are connected by 4*200 Gbps RDMA network based on RoCEv2 with rail-optimized topology.
The overall experiments use up to 1296 GPUs, and the \revisionsigcomm{ablation study} utilizes up to 96 GPUs.
We use PyTorch 2.1.2 and NVIDIA CUDA 12.2.
\revisionsigcomm{
As for the modality, we focus on images and texts.
We emphasize that \sysname is also compatible with other modalities.
Our evaluation focuses on images (i.e., vision). This choice is motivated
by the prevalence of vision as a common non-textual modality in multimodal LLMs
and the increasing prominence of capable vision-based MLLMs (e.g., Gemini~\cite{gemini} and Qwen2.5-VL~\cite{bai2025qwen2}).
}

\parabf{Models.}
For the LLM backbone, we choose the representative LLM architecture, Llama3~\cite{llama},
which is widely used in both academia and industry.
Table~\ref{eval:tab:datasets} lists the detailed model configurations.
For encoder and generator, we use ViT-Huge~\cite{vit-huge} (0.63B) and SD 2.1~\cite{sd} (1B) respectively.
\revisionsigcomm{
The encoder and generator in our experiments are representative of state-of-the-art MLLM.
The encoder (i.e., ViT-Huge) in our experiment aligns with the encoders of Qwen2.5-VL and Seed1.5-VL
in terms of both model architecture and size.
The generator (i.e., SD) is also adopted in many MLLMs (e.g., Step1X-Edit~\cite{liu2025step1x} and EMU~\cite{sun2023emu}) with generation capabilities.
}
The three LLM backbones (i.e., Llama3-7B, Llama3-13B, and Llama3-70B) are paired with ViT-Huge and SD
to form multimodal LLMs denoted as MLLM-9B, MLLM-15B, and MLLM-72B.
For the large multimodal LLM (i.e., MLLM-72B), we use high image resolution (i.e., 1024$\times$1024) for
generation since the large LLM is able to process more context information.
For small models, we use low image resolution (i.e., 512$\times$512).

\begin{table}[t!]
    \centering
    \resizebox{\linewidth}{!} {
    \begin{tabular}{cccccc}
        \toprule
        \multirow{2}{*}{\textbf{Models}} & \textbf{\# of} & \textbf{Hidden} &
        \textbf{FFN Hidden} & \textbf{\# of} & \textbf{\# of} \\
         & \textbf{Layers} & \textbf{Size} & \textbf{Size} & \textbf{Heads} & \textbf{Groups} \\
        \midrule
        Llama3-7B & 32 & 4096 & 11008 & 32 & 32\\
        Llama3-13B & 40 & 5120 & 13824 & 40 & 40\\
        Llama3-70B & 80 & 8192 & 28672 & 64 & 8\\
        \bottomrule
    \end{tabular}
    }
    \vspace{-0.0in}
    \caption{LLM backbone configurations.}
    \vspace{-0.4in}
    \label{eval:tab:datasets}
\end{table}

\parabf{Datasets.}
For our experiments, we use
the representative open-source dataset, LAION-400M~\cite{schuhmann2021laion}. We generate training data by interleaving
the image and text subsequences, forming input sequences up to 8192 tokens long.
This dataset is also employed in our production multimodal LLM training.
As detailed in \S\ref{sec:background:data}, each training sample includes a varying number of image tokens
and text tokens, which introduces data heterogeneity in multimodal LLM training.

\parabf{Metrics.}
We use Model FLOPs Utilization (MFU) as the primary metric.
MFU measures the percentage of GPU FLOPs that
are effectively utilized during training. We also use the training throughput
to evaluate the training speed of \sysname.
Since \sysname and Megatron-LM may utilize different numbers of GPUs due to
varying model orchestration strategies, we also indicate the number of GPUs used in each experiment.

\parabf{Baselines.}
\revision{
We focus on training multimodal LLMs rather than traditional multimodal models like CLIP or LiT.
DistMM~\cite{huang2024distmm} efficiently handles traditional multimodal training but lacks support
for pipelines involving encoders, LLM backbones, and generators in multimodal LLMs.
Multimodal LLMs combine multimodal models and LLMs to understand and generate high-quality
multimodal data auto-regressively.
\revisionsigcomm{
Megatron-LM is selected as the primary baseline due to its established role as a highly optimized,
open-source framework for LLM training that also provides supports for multimodal models.
}
For ablation study of model orchestration, we integrate DistMM's model orchestration strategy
(resource allocation by model size and FLOPs) into \sysname,
naming this baseline DistMM$^*$. Note that DistMM does not support
multimodal LLM training natively. DistMM$^*$ only uses its orchestration strategy,
with all other techniques from \sysname.
}


\subsection{Overall Performance}
\label{sec:evaluation:overall}

\begin{figure}[t!]
    \centering
    \includegraphics[width=0.76\linewidth]{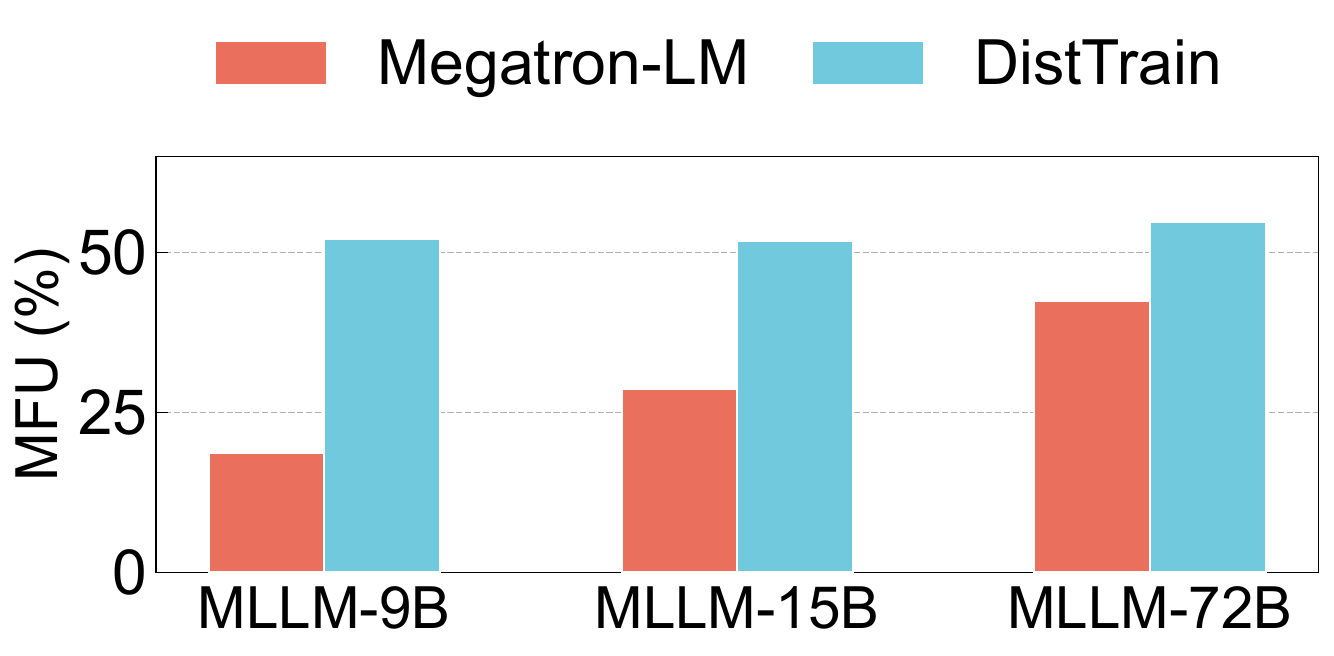}
    \vspace*{-0.15in}
    \caption{The overall MFU of \sysname and Megatron-LM.}
    \vspace*{-0.1in}
    \label{fig:evaluation:overall_mfu}
\end{figure}

We first compare the overall performance of \sysname against Megatron-LM
on a large GPU cluster (up to 1296 GPUs). We retrofit Megatron-LM to support multimodal
LLM training by integrating modality encoder and generator into the training pipeline.
Megatron-LM employs monolithic model orchestration as described in \S\ref{sec:background:bg}.
In Megatron-LM, we set the PP size of the LLM backbone to 1, 2, and 10 for Llama3-7B, Llama3-13B, and Llama3-70B, respectively.
PP size is set to 1 for modality encoder and generator.
We set TP size to 8, as each node consists of 8 GPUs connected by high-bandwidth NVLink. As for \sysname, the parallelism strategy
is determined by disaggregated model orchestration.
In our experiments,
one GPU is able to facilitate training ViT and SD.
We replicate the modality encoder and generator across the GPUs
within the TP group to process different images, whereas TP itself is not used.
We set the global batch size to 1920.

\begin{figure}[t!]
    \centering
    \includegraphics[width=0.76\linewidth]{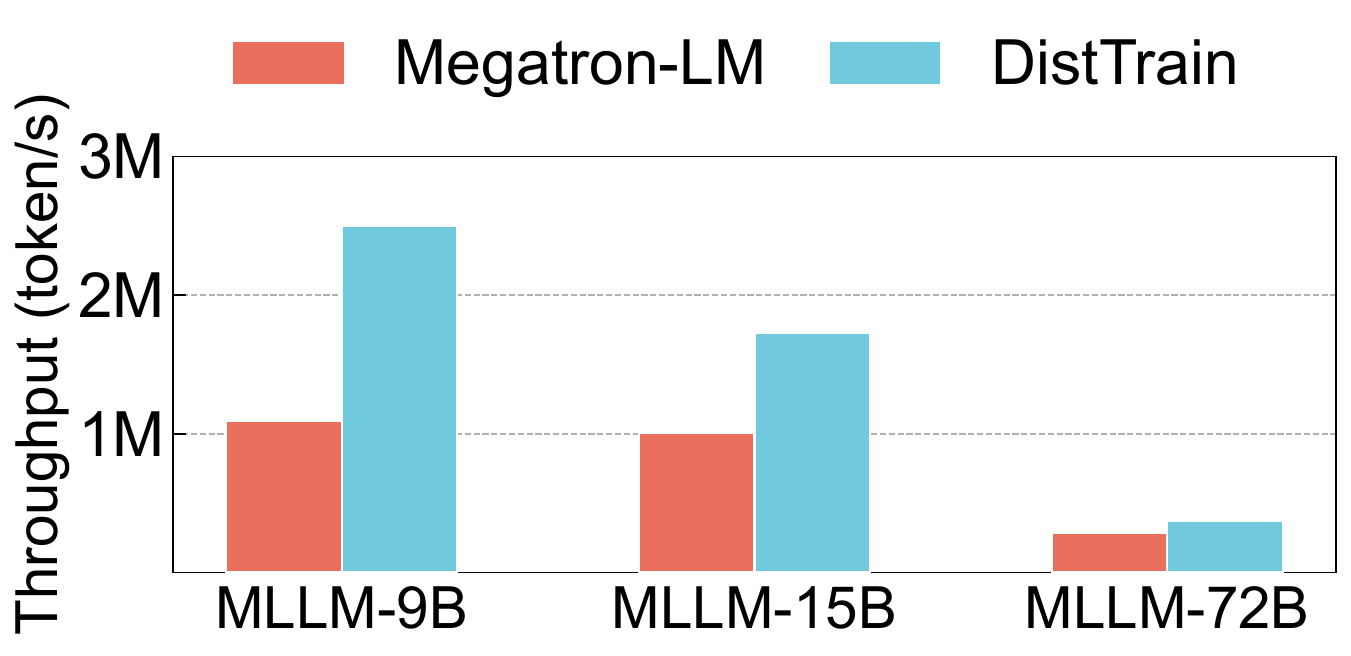}
    \vspace*{-0.15in}
    \caption{The overall throughput of \sysname and Megatron-LM.}
    \vspace*{-0.1in}
    \label{fig:evaluation:overall_throughput}
\end{figure}

The experimental results are shown in Figure~\ref{fig:evaluation:overall_mfu} and Figure~\ref{fig:evaluation:overall_throughput}.
Figure~\ref{fig:evaluation:overall_mfu} shows the MFU. Figure~\ref{fig:evaluation:overall_throughput} shows the training throughput.
Due to the different model orchestration strategies of \sysname and Megatron-LM,
\sysname uses 1056, 1216, and 1176 GPUs for MLLM-9B, MLLM-15B, and MLLM-72B, respectively.
Megatron-LM uses 1296, 1280, and 1152 GPUs, respectively.
\revisionsigcomm{
Notably, within a total budget of 1296 GPUs, \sysname intentionally
allocates fewer resources in some cases because adding more GPUs yields
no further improvements in training throughput.
This resource efficiency frees the remaining GPUs for concurrent tasks such as fine-tuning or inference.
}
We summarize the findings as follows.
\begin{itemize}[leftmargin=*]
    \item As shown in Figure~\ref{fig:evaluation:overall_mfu}, \sysname achieves 51.8\%-54.7\% MFU
    in large-scale multimodal LLM training. This performance
    closely approximates that of state-of-the-art unimodal (i.e., text) LLM training~\cite{jiang2024megascale},
    which demonstrates the
    effectiveness of \sysname in addressing the model and data heterogeneity in
    multimodal LLM training.

    \item \sysname significantly outperforms Megatron-LM, delivering 1.7--2.8$\times$ the
    MFU and 1.7--2.2$\times$ the training throughput when training MLLM-9B and MLLM-15B with a similar number of GPUs.
    These performance gains largely stem from \sysname's disaggregated model orchestration.
    Megatron-LM's monolithic strategy often leads to GPU underutilization,
    since it assigns too many GPUs to the modality encoder and generator. In contrast,
    \sysname adaptively adjusts model orchestration based on specific model and data demands.
    Additionally, \sysname's disaggregated data preprocessing technique further improves efficiency.

    \item In the MLLM-72B training scenario, \sysname outperforms
    Megatron-LM by 1.2$\times$ on MFU and 1.3$\times$ on training throughput with a similar number of GPUs.
    The high image resolution increases the execution time of the multimodal module, which introduces pipeline bubbles in LLM backbone.
    \sysname addresses this by allocating additional GPUs to these modules to balance the pipeline.
    The disaggregated data preprocessing strategy continues to mitigate data heterogeneity,
    thereby increasing training efficiency.
\end{itemize}

\subsection{Ablation Study}
\label{sec:evaluation:deep}
In this subsection, we perform microbenchmarks to evaluate the effectiveness
of each \sysname's technique and utilize up to 96 GPUs.
\revisionsigcomm{
We reduce the cluster scale due to our limited budget.
}
We set the global batch size of training to 128, 64, and 40 for MLLM-9B, MLLM-15B, and MLLM-72B, respectively.

\begin{figure}[t!]
    \centering
    \includegraphics[width=1\linewidth]{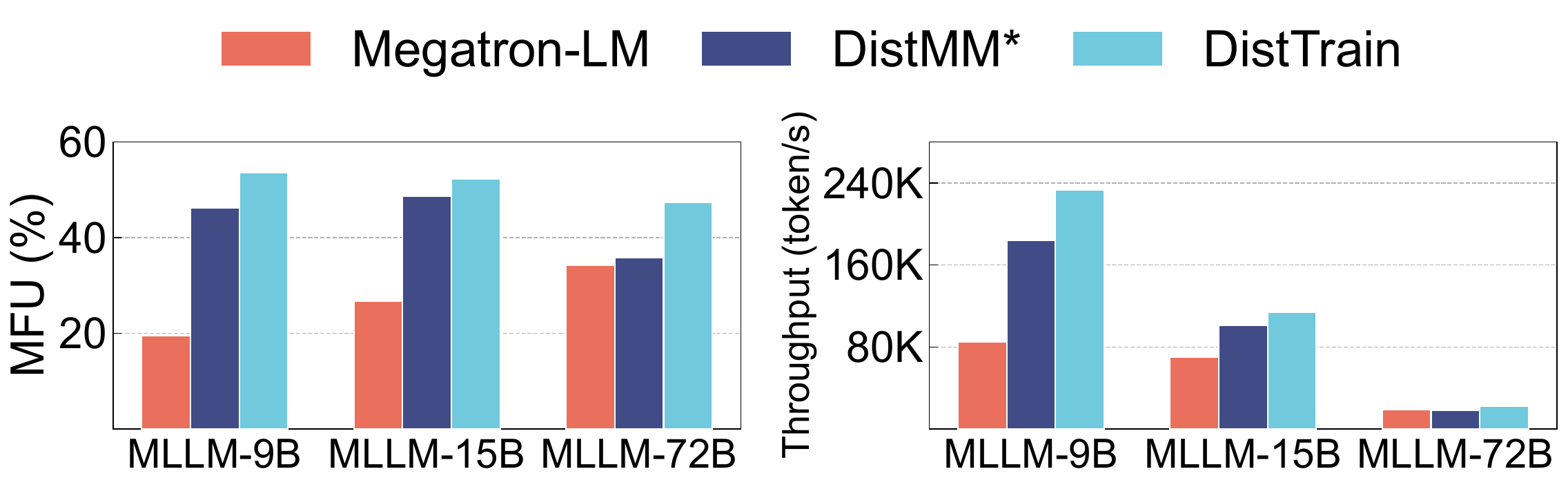}
    \small{\hspace*{0.3in}{(a) MFU.}\hspace*{\dimexpr\linewidth/3\relax}{(b) Throughput.}}
    \vspace*{-0.1in}
    \caption{Disaggregated model orchestration.}
    \vspace*{-0.0in}
    \label{fig:evaluation:resource_allocation}
\end{figure}

\parabf{Disaggregated model orchestration.}
\revision{
We evaluate the MFU and training throughput using the model orchestration strategies of \sysname, Megatron-LM, and DistMM$^*$.
DistMM$^*$ allocates GPUs according to the computation demands (flops) of each module.
\sysname uses 96, 80, and 82 GPUs for MLLM-9B, MLLM-15B, and MLLM-72B, respectively.
Megatron-LM uses 96, 96, and 96 GPUs, while DistMM$^*$ uses 88, 76, and 90 GPUs.
The experimental results are shown in Figure~\ref{fig:evaluation:resource_allocation}.
\sysname consistently outperforms the baseline strategies, achieving 1.3--2.7$\times$ higher MFU and 1.4--2.7$\times$ higher
training throughput.
Although DistMM$^*$ outperforms Megatron-LM,
it still lags behind \sysname since it neglects the intricate performance model
(\S\ref{sec:design:resource:form}) of multimodal LLM training. \sysname's disaggregated model orchestration
optimally balances computational loads across the three modules
and achieves high resource utilization.
In production training, the running time of disaggregated model orchestration
is less than one second, negligible compared to the days required for full training.
}

\revisionsigcomm{
We also evaluate the running time of \sysname's disaggregated model orchestration
to determine the optimal resource and parallelism configurations under different training settings,
as detailed in Table~\ref{eval:tab:resource:overhead}. The algorithm completes
in under one second. The overhead is negligible compared to the days or even weeks
required for overall training.
}

\begin{figure}[t!]
    \centering
    \includegraphics[width=1\linewidth]{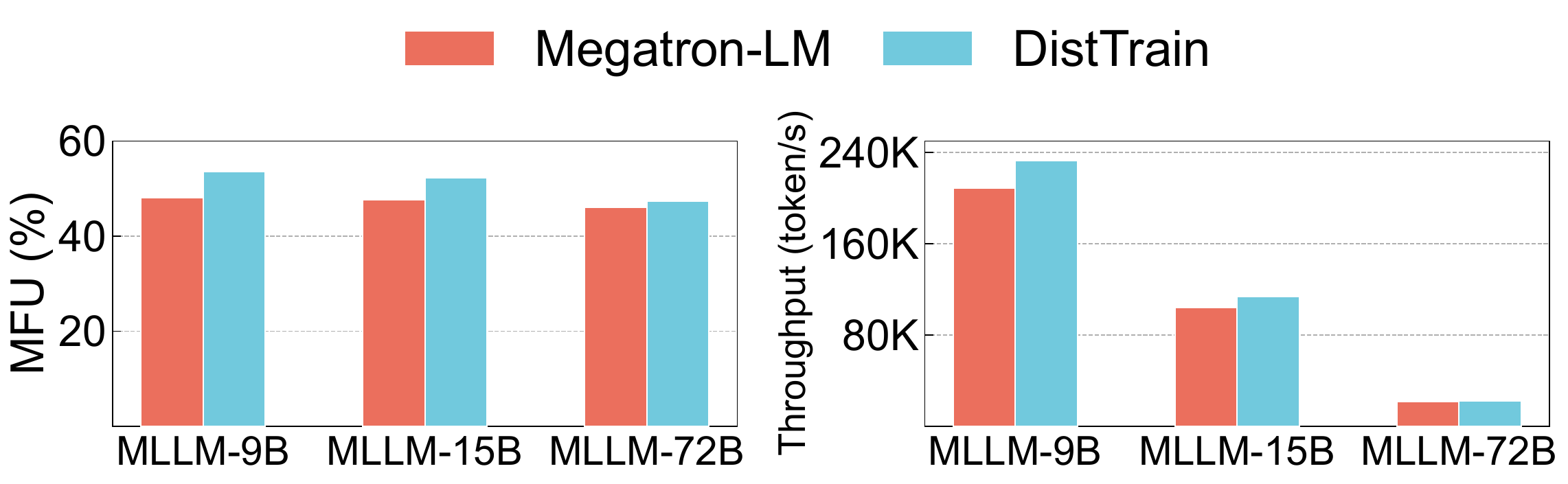}
    \small{\hspace*{0.3in}{(a) MFU.}\hspace*{\dimexpr\linewidth/3\relax}{(b) Throughput.}}
    \vspace*{-0.1in}
    \caption{Disaggregated data preprocessing.}
    \vspace*{-0.05in}
    \label{fig:evaluation:reordering}
\end{figure}

\parabf{Disaggregated data preprocessing.}
We evaluate the effectiveness of \sysname's disaggregated data preprocessing by comparing it against Megatron-LM's data preprocessing,
while keeping other components the same.
Megatron-LM's data preprocessing uses random ordering for training data.
The effectiveness is gauged through metrics such as MFU and training throughput. We use the optimal resource allocation
and parallelism strategy decided by \sysname's disaggregated model orchestration.
Given that the model orchestration strategy remains unchanged, we do not indicate the number of GPUs.
The experimental settings are the same as those in the experiment to evaluate disaggregated model orchestration.
The results are shown in Figure~\ref{fig:evaluation:reordering}.
\sysname consistently outperforms the baseline, achieving 1.03-1.11$\times$ higher MFU and
throughput. The performance gap becomes more pronounced as the model size decreases.
This is because the smaller model size leads to a higher data parallelism (DP) size, which causes more
intra-microbatch heterogeneity. In essence, \sysname's two-level data reordering effectively mitigates
data heterogeneity and enhances training efficiency.
We do not measure the running time of the reordering algorithm as it operates on dedicated CPU nodes asynchronously.
The data preprocessing overhead with and without disaggregation is evaluated in \S\ref{sec:evaluation:case}.

\begin{table}[t!]
    \centering
    \resizebox{0.66\linewidth}{!} {
    \begin{tabular}{cccc}
        \toprule
        \textbf{Model} & \textbf{\# of} & \textbf{Global} &
        \textbf{Algorithm} \\
         & \textbf{GPUs} & \textbf{Batch Size} & \textbf{Overhead} \\
        \midrule
        MLLM-72B & 1296 & 1920 &  922ms \\
        MLLM-72B & 648 & 960 &  641ms \\
        MLLM-72B & 324 & 480 &  441ms \\
        MLLM-72B & 112 & 240 &  133ms \\
        \bottomrule
    \end{tabular}
    }
    \vspace{0.1in}
    \caption{\revisionsigcomm{Overhead of disaggregated model orchestration.}}
    \vspace{-0.1in}
    \label{eval:tab:resource:overhead}
\end{table}

\subsection{Case Study}
\label{sec:evaluation:case}
In this subsection, we first evaluate the overhead of data preprocessing.
We then evaluate \sysname under different frozen training settings.

\begin{figure}[t!]
    \centering
    \includegraphics[width=0.75\linewidth]{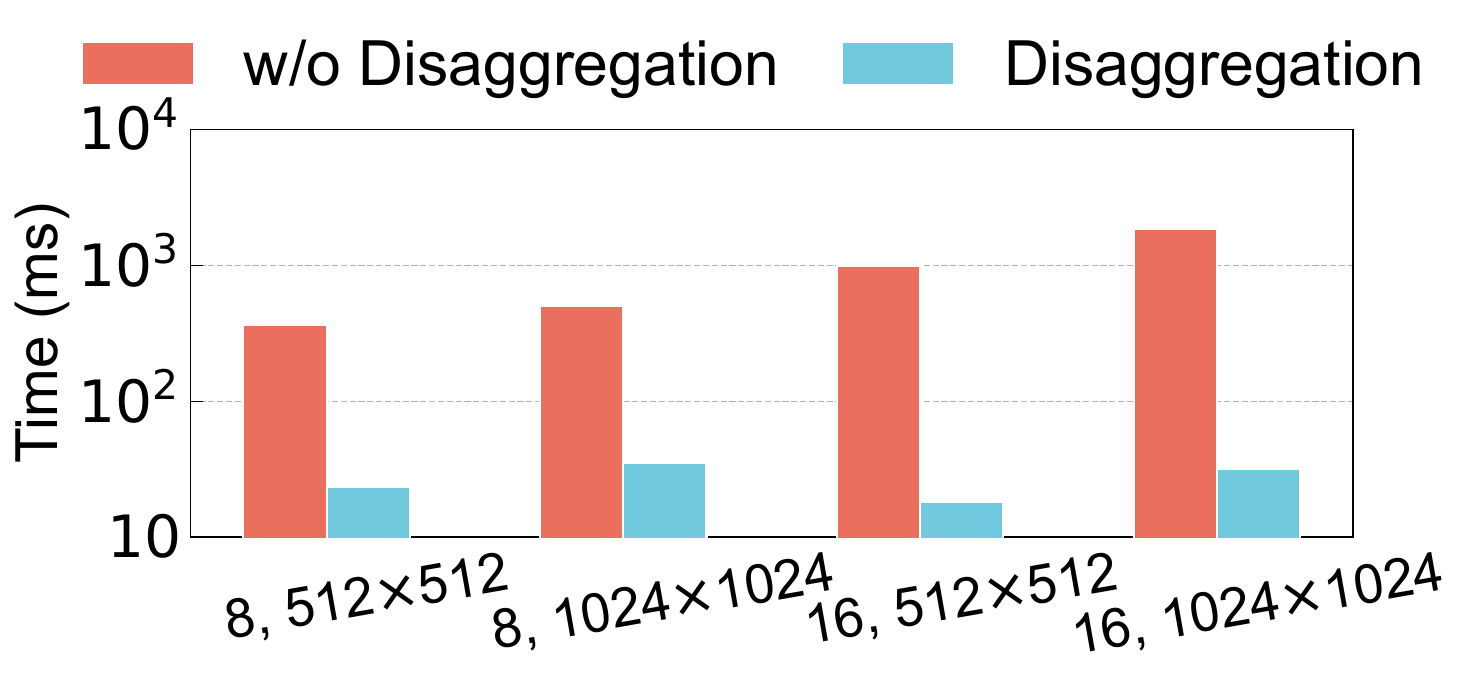}
    \vspace*{-0.1in}
    \caption{Overhead of data preprocessing.}
    \vspace*{-0.2in}
    \label{fig:app:sys_opt_async_load}
\end{figure}

\parabf{Overhead of data preprocessing.}
We conduct an experiment to evaluate the overhead of data preprocessing,
including decompression and reordering. Setting the DP size to one, we measure
the average data preprocessing time per iteration on the GPU training side. We
then compare data preprocessing time with and without disaggregated data
preprocessing and use varying numbers of images and different image resolutions
for one training iteration. The results, depicted in
Figure~\ref{fig:app:sys_opt_async_load}, indicate disaggregating data
preprocessing from training reduces preprocessing time from seconds to
milliseconds. The first parameter in the x-axis represents the number of images,
while the second parameter denotes the image resolution. In production training
(\S\ref{sec:evaluation:overall}), iteration times range from seconds to tens of
seconds. Without disaggregation, preprocessing overhead is in seconds,
significantly interferes with training. With disaggregated data preprocessing,
the overhead reduces to milliseconds, which is negligible relative to total
iteration time.

\begin{figure*}[t!]
    \centering
    \includegraphics[width=1\linewidth]{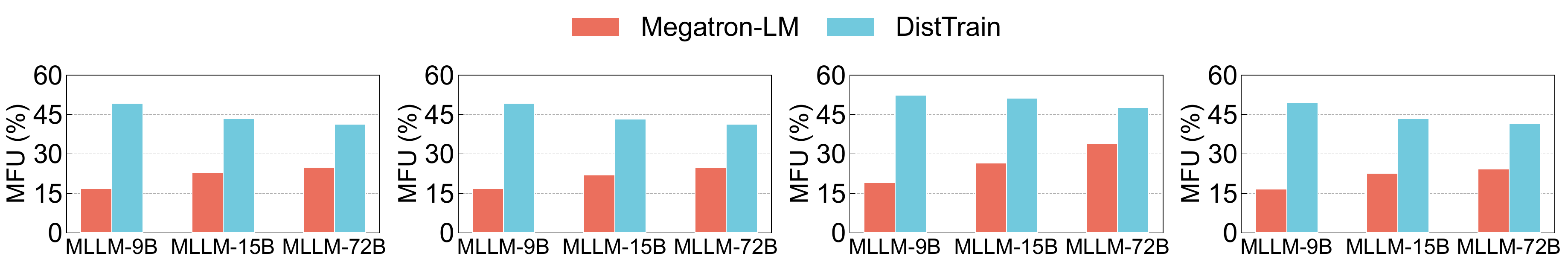}
    \small{\hspace*{0.35in}{(a) All modules freezing.}\hspace*{\dimexpr\linewidth/16\relax}{(b) Encoder-only training.}\hspace*{\dimexpr\linewidth/16\relax}{(c) LLM-only training.}\hspace*{\dimexpr\linewidth/16\relax}{(d) Generator-only training.}}
    \vspace*{-0.1in}
    \caption{MFU under frozen training setting.}
    \vspace*{-0.1in}
    \label{fig:evaluation:freeze_mfu}
\end{figure*}

\begin{figure*}[t!]
    \centering
    \includegraphics[width=1\linewidth]{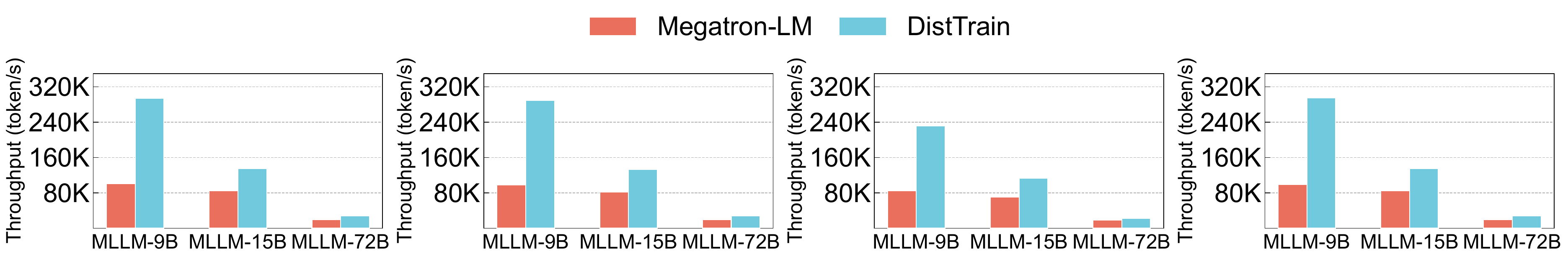}
    \small{\hspace*{0.35in}{(a) All modules freezing.}\hspace*{\dimexpr\linewidth/16\relax}{(b) Encoder-only training.}\hspace*{\dimexpr\linewidth/16\relax}{(c) LLM-only training.}\hspace*{\dimexpr\linewidth/16\relax}{(d) Generator-only training.}}
    \vspace*{-0.1in}
    \caption{Throughput under frozen training setting.}
    \vspace*{-0.1in}
    \label{fig:evaluation:freeze_throughput}
\end{figure*}

\parabf{Frozen training.}
\revision{
In real production training, some modules in multimodal LLM are frozen to stabilize training
loss and enhance model effectiveness during different training phases~\cite{yin2023survey}.
}
We conduct a frozen training experiment under four specific training settings: complete module freezing (i.e., training projectors only),
encoder-only training, LLM-only training, and generator-only training.
In these scenarios, frozen modules neither compute weight gradients in backward pass nor update weights.
The frozen modules still perform computations in forward pass.
All other experimental setup aligns with those detailed in \S\ref{sec:evaluation:deep}.
\sysname uses 96, 80, and 82 GPUs for MLLM-9B, MLLM-15B, and MLLM-72B under all three frozen settings.
Megatron-LM uses 96, 96, and 96 GPUs for MLLM-9B, MLLM-15B, and MLLM-72B under all three frozen settings.
The experimental results are in Figure~\ref{fig:evaluation:freeze_mfu} and Figure~\ref{fig:evaluation:freeze_throughput}.
\sysname consistently outperforms Megatron-LM across all frozen training configurations, achieving 1.4--2.9$\times$ higher MFU and
1.2--2.9$\times$ higher training throughput.
This pronounced performance gap underscores the challenges posed by Megatron-LM's monolithic model orchestration in more complex training environments.
In contrast, \sysname adaptively adjusts model orchestration based on training settings and
achieves high resource utilization.
\section{Discussion}
\label{sec:design:discussion}

\paraf{Parallelism strategies.}
\revisionsigcomm{
Many studies~\cite{zheng2022alpa, wang2019supporting, jia2019beyond} optimize parallelism strategies to accelerate deep learning model training. However, they fall short
of holistically considering data, pipeline, and tensor parallelism in multimodal LLM training, due to the large search spaces in large clusters.
This limitation results in inefficient and sub-optimal parallelism strategy.
Additionally, these methods generally assume homogeneity of training data---all training samples consume the same amount of computation.
This assumption does not hold for multimodal LLMs where training samples of different modalities vary significantly in volume and computation demand.
\sysname leverages the specific training pattern of multimodal LLMs, and formulates
a model orchestration problem for multimodal LLMs
that integrates tensor, pipeline, and data parallelism simultaneously.
Besides, \sysname leverages disaggregated data reordering to address data heterogeneity.
}

\parabf{Sequence and expert parallelism.}
\revisionsigcomm{
Sequence parallelism (SP)~\cite{li2021sequence} is designed to partition the training sequence into multiple subsequences for parallel training.
It addresses the challenges of processing long sequences in LLMs. Expert parallelism (EP)~\cite{liu2023janus}, specifically devised for mixture-of-experts (MoE) LLMs~\cite{du2022glam},
enables parallel training of multiple feedforward network (FFN) experts.
These parallelism strategies are orthogonal to multimodal LLM training.
In \sysname, both SP and EP are integrated into the LLM backbone training.
}

\parabf{Heterogeneous hardware.}
\revisionsigcomm{
By disaggregating three modules (i.e., the modality encoder, LLM backbone, and generator) for multimodal LLM training,
\sysname supports using heterogeneous hardware for different modules to
achieve varius goals like
improving performance, reducing cost and increasing energy efficiency.
This strategy assigns distinct computational workloads to the most appropriate hardware.
For instance, ViT encoder is significantly less compute-intensive than LLM backbone.
Consequently, we can place ViT encoder on more economical GPUs
(e.g., NVIDIA L20).
Such benefits are demonstrated in other LLM systems such as MegaScale-Infer~\cite{zhu2025megascale} and Splitwise~\cite{patel2024splitwise}.
}

\parabf{Disaggregated data preprocessing.}
The principal technique in \sysname's disaggregated data preprocessing is data
reordering that addresses training stragglers caused by data heterogeneity.
Although data reordering does not
inherently necessitate disaggregation, we adopt disaggregation to mitigate the
preprocessing overhead and improve system elasticity (e.g., allocating any
arbitrary number of CPUs for preprocessing tasks). This approach is analogous to
resource disaggregation in other  systems scenarios, which aims to improve
system elasticity and efficiency~\cite{shan2018legoos, wang2022memliner}.
\section{Related Work}
\label{sec:related}

\paraf{LLM training.}
Many efforts have been made to optimize LLM training.
For LLM pre-training, Megatron-LM~\cite{shoeybi2019megatron} and DeepSpeed-Megatron~\cite{smith2022using} propose custom 3D-parallelism strategies.
DeepSpeed-ZeRO~\cite{rajbhandari2020zero} and Pytorch-FSDP~\cite{zhao2023pytorch} reduce redundant memory consumption in data parallelism.
A set of works~\cite{chen2024centauri, jiang2024megascale, bytedance_flux, TE, asplos2022overlap} overlap the communication and computation operators in LLM training.
These systems optimizations of LLM training are orthogonal to \sysname. They do not address
model and data heterogeneity in multimodal LLM training.
\sysname integrates several of these optimizations in training the LLM backbone.

\parabf{Traditional multimodal training.}
\revision{Traditional} multimodal models (e.g., CLIP~\cite{radford2021learning} and LiT~\cite{zhai2022lit}) have been widely studied in recent years.
Many systems optimizations have been proposed to train such multimodal models efficiently.
\revision{
DistMM~\cite{huang2024distmm} optimizes multiple parallel encoders training
for contrastive learning.}
Yet, these advancements primarily enhance traditional multimodal models (e.g., CLIP~\cite{radford2021learning} and LiT~\cite{zhai2022lit}), which do not integrate LLM
for understanding and generation (e.g., Flamingo~\cite{alayrac2022flamingo} and LlaVa~\cite{liu2024visual}). They do not support encoder, LLM backbone and generator training pipeline
in multimodal LLMs.
Therefore, the state-of-the-art solution is to leverage Megatron-LM to train multimodal LLMs at large scale.

\section{Conclusion}
\label{sec:conclusion}

We present \sysname, a disaggregated multimodal LLM training system. We identify
the key challenges in training multimodal LLMs, i.e., model heterogeneity and
data heterogeneity. \sysname introduces disaggregated model orchestration to
address model heterogeneity and disaggregated data preprocessing to address data
heterogeneity. We evaluate \sysname on a large production cluster with thousands
of GPUs and show that it achieves 54.7\% MFU and outperforms Megatron-LM by up
to 2.2$\times$ on training throughput.

\parait{This work does not raise any ethical issues.}

\parabf{Acknowledgments.} 
We sincerely thank the anonymous reviewers for their valuable feedback on this
paper. Xin Jin is the corresponding author. Zili Zhang, Yinmin Zhong,
and Xin Jin are also with the Key Laboratory of High
Confidence Software Technologies (Peking University), Ministry of Education. 

This work was supported in part by the National Key Research and Development Program
of China under Grant 2022YFB4500700, the Scientific Research Innovation Capability
Support Project for Young Faculty under Grant ZYGXQNJSKYCXNLZCXM-I1,
the Fundamental Research Funds for the Central Universities, Peking University,
and the National Natural Science Foundation of China under Grant 62172008.

\def\UrlBreaks{\do\/\do-}
\bibliographystyle{ACM-Reference-Format}
\bibliography{xin}

\clearpage

\appendix
\parait{Appendices are supporting material that has not been peer-reviewed.}
\section{Appendix}
\label{appendix}
\pagestyle{plain}

\subsection{Mitigating TP overhead with StepCCL}
\label{appendix:ccl}

\begin{figure}[h]
    \centering
    \includegraphics[width=.72\linewidth]{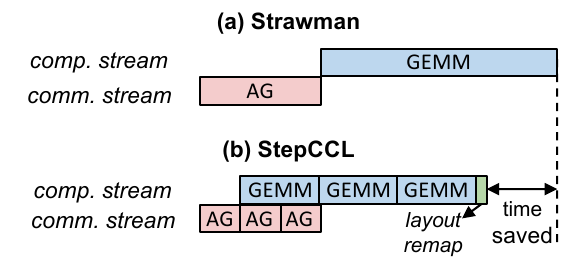}
    \vspace{-0.1in}
    \caption{Overlapping communication and computation.}
    \vspace{-0.15in}
    \label{fig:design:ccl_example}
\end{figure}

Tensor Parallelism (TP) is commonly adopted to facilitate training large Transformer-based models with multiple GPUs connected with high bandwidth (e.g., NVLinks). Specifically, TP divides the linear layers, into smaller sub-modules, which are then distributed across the GPUs. After parallel computation, all GPUs perform collective communication to aggregate data and produce identical results. The TP communication overhead severely degrades overall performance.

We implement the communication overlap with an in-house collective communication library called StepCCL to reduce the TP overhead. StepCCL is a PyTorch custom plugin that performs cross-GPU collective communication, including allgather, reduce-scatter, and allreduce, which is similar to NCCL. However, NCCL occupies several CUDA Streaming Multiprocessors (SMs) for executing its communication kernel and is known to harm the performance of its concurrent GEMM (matrix multiplication) ~\cite{isca2021overlap}. To solve this, StepCCL leverages the DMA engine directly to perform data transmission without using any SM at all. This enables StepCCL and GEMM to run simultaneously on a GPU without slowing down each other. This cornerstone facilitates our subsequent development of communication overlap. 

\begin{figure}[t]
    \centering
    \includegraphics[width=\linewidth]{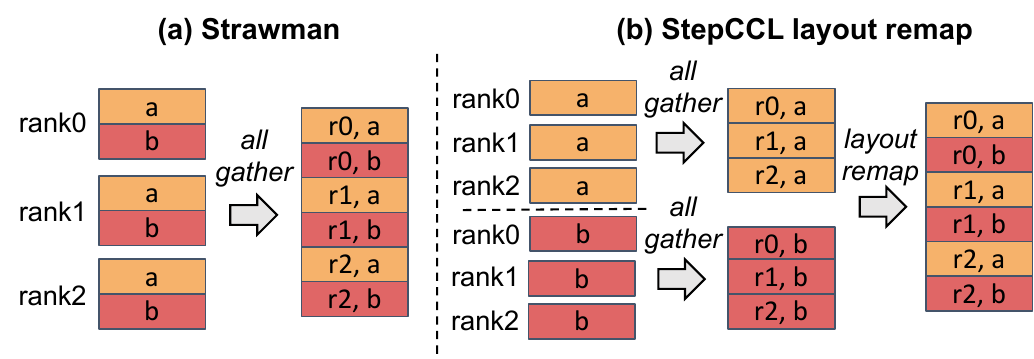}
    \vspace{-0.25in}
    \caption{Layout remap.}
    \vspace{-0.1in}
    \label{fig:design:ccl_layout}
\end{figure}

Figure~\ref{fig:design:ccl_example} shows an example of how StepCCL works in overlapping the allgather (AG) operation with GEMM. We start by decomposing the single GEMM and the corresponding communication into several smaller pairs. Each small communication operation starts sequentially on a communication stream, with its paired GEMM executed on the default computation stream. The communication overhead is fully hidden except for the first allgather.\footnote{If the number of allgather/GEMM is large enough, the only allgather in the critical path should have negligible overhead. But dividing a large GEMM into finer granularity sometimes could lead to overall slowdown. In practice, the number is actually configurable.} 
After all GEMMs finish, we perform an extra layout remapping operation (usually with negligible overhead) to ensure identical results with the baseline. Figure~\ref{fig:design:ccl_layout} describes the details of the layout remap process. In some rare cases during backward propagation, we find the remap overhead is high due to certain model dimensions. To mitigate this, we further overlap the remap with the computation of the weight gradients, so eventually we nearly get the full performance gain of the communication overlap.

Finally, although the overlap idea is also studied in many related works~\cite{bytedance_flux, TE, asplos2022overlap}, we highlight the key differences of ours.
Unlike prior work that fuses GEMM with TP communication into one CUDA kernel~\cite{bytedance_flux, TE}, we choose a modular design and do not use fusion for more flexibility. For example, when TP communication is longer than GEMM, fusing them cannot fully hide the communication overhead. However, with the modular design, we are able to hide the communication with other modules without dependency (e.g., in cross-attention), which is not possible with the fused implementation. This enables broader adoption of StepCCL in many other scenarios.

\begin{figure}[t!]
    \centering
    \includegraphics[width=1\linewidth]{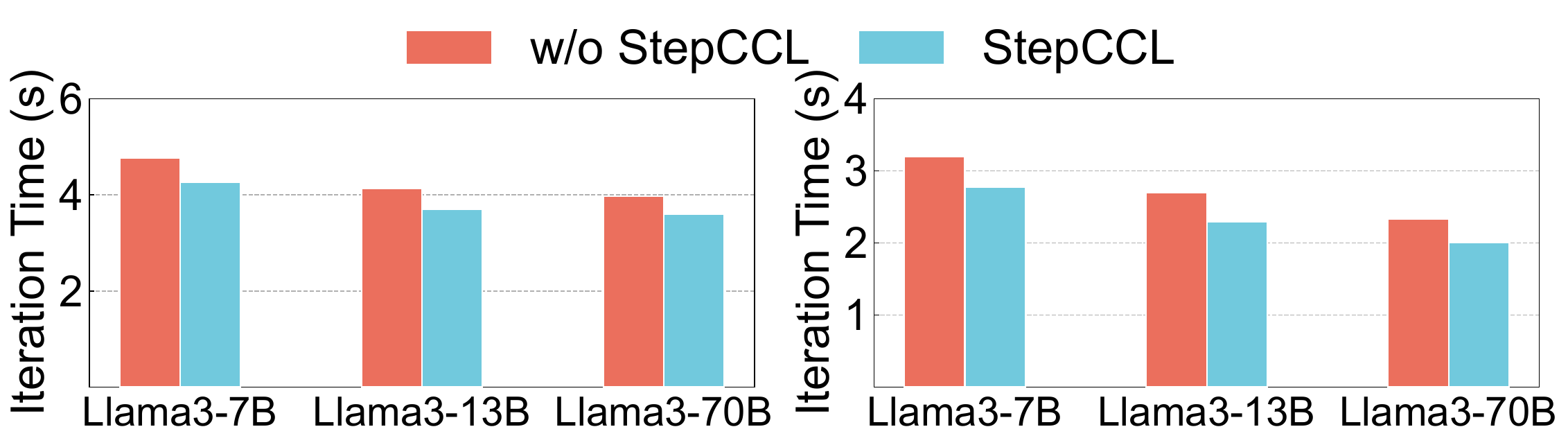}
    \small{\hspace*{0.3in}{(a) TP=4.}\hspace*{\dimexpr\linewidth/3\relax}{(b) TP=8.}}
    \vspace*{-0.1in}
    \caption{Overlapping the TP communication with computation.}
    \vspace*{-0.15in}
    \label{fig:app:sys_opt_tp_overlap}
\end{figure}

\parabf{Evaluation.}
To evaluate the effectiveness of StepCCL in mitigating the TP overhead,
we conduct an experiment that measures the iteration time of the LLM backbone with
training of one single PP stage (i.e., one minimal TP group) under various TP sizes.
We compare the iteration time with and without StepCCL enabled.
The results are shown in Figure~\ref{fig:app:sys_opt_tp_overlap}.
StepCCL significantly reduces the iteration time by overlapping the TP communication with computation.
It outperforms the baseline by 1.1-1.12$\times$ when the TP size is 4 and 1.15-1.17$\times$ when the TP size is 8.
The gains are more pronounced at large TP size, where communication overhead is more substantial.
These findings confirm that StepCCL effectively mitigates TP overhead.

\end{sloppypar}
\end{document}